\documentclass[authorversion]{acmart}
\AtBeginDocument{%
  }

\usepackage{color}
\usepackage{xcolor}
\usepackage{booktabs} 
\usepackage{multirow}
\usepackage{amsmath}
\usepackage{amsthm} 
\usepackage{url}
\usepackage{wrapfig}
\usepackage{xspace}
\usepackage{pifont} 
\usepackage{cleveref}
\usepackage{wrapfig}

\newcommand{\x}{\mathbf{x}}

\newcommand{\hOneMain}{\textbf{H}$_\textbf{1a}$\xspace}
\newcommand{\hOneStakes}{\textbf{H}$_\textbf{1b}$\xspace}
\newcommand{\hOneFraming}{\textbf{H}$_\textbf{1c}$\xspace}
\newcommand{\hOneUnc}{\textbf{H}$_\textbf{1d}$\xspace}
\newcommand{\hTwoMain}{\textbf{H}$_\textbf{2a}$\xspace}
\newcommand{\hTwoStakes}{\textbf{H}$_\textbf{2b}$\xspace}
\newcommand{\hTwoFraming}{\textbf{H}$_\textbf{2c}$\xspace}
\newcommand{\hTwoUnc}{\textbf{H}$_\textbf{2d}$\xspace}

\newcommand{\explLikert}{\textbf{E1}\xspace}
\newcommand{\explDuration}{\textbf{E2}\xspace}
\newcommand{\explTrends}{\textbf{E3}\xspace}
\newcommand{\explCost}{\textbf{E4}\xspace}
\newcommand{\explQual}{\textbf{E5}\xspace}

\newcommand{\extracost}{\texttt{extra cost}\xspace}
\newcommand{\nocost}{\texttt{no extra cost}\xspace}

\newcommand{\implication}[1]{
\vspace{2mm}\noindent\fbox{
    \parbox{0.96\linewidth}{%
    \emph{Implication:} #1
    }
}
\vspace{2mm}
}

\copyrightyear{2025}
\acmYear{2025}
\setcopyright{cc}
\setcctype{by}
\acmConference[CHI '25]{CHI Conference on Human Factors in Computing Systems}{April 26-May 1, 2025}{Yokohama, Japan}
\acmBooktitle{CHI Conference on Human Factors in Computing Systems (CHI '25), April 26-May 1, 2025, Yokohama, Japan}\acmDOI{10.1145/3706598.3713524}
\acmISBN{979-8-4007-1394-1/25/04}

\begin{document}

\title{Perceptions of the Fairness Impacts of Multiplicity in Machine Learning}

\author{Anna P. Meyer}
\email{apmeyer4@wisc.edu}
\orcid{0009-0008-9763-5585}
\affiliation{%
  \institution{University of Wisconsin - Madison}
  \city{Madison}
  \state{Wisconsin}
  \country{USA}
}
\authornote{Correspondence to \href{mailto:apmeyer4@wisc.edu}{apmeyer4@wisc.edu}   }

\author{Yea-Seul Kim}
\email{yeaseul.kim@gmail.com}
\orcid{0000-0003-1854-1537}
\affiliation{%
  \institution{Apple}
  \city{Boulder}
  \state{Colorado}
  \country{USA}
}

\author{Aws Albarghouthi}
\email{aws@cs.wisc.edu}
\orcid{0000-0003-4577-175X}
\affiliation{%
  \institution{University of Wisconsin - Madison}
  \city{Madison}
  \state{Wisconsin}
  \country{USA}
}

\author{Loris D'Antoni}
\email{ldantoni@ucsd.edu}
\orcid{0000-0001-9625-4037}
\affiliation{%
  \institution{University of California - San Diego}
  \city{San Diego}
  \state{California}
  \country{USA}
}

\renewcommand{\shortauthors}{Meyer et al.}

\begin{abstract}
    Machine learning (ML) is increasingly used in high-stakes settings, yet  \emph{multiplicity} -- the existence of multiple good models -- means that some predictions are essentially arbitrary. 
ML researchers and philosophers posit that multiplicity poses a fairness risk, but no studies have investigated 
    whether stakeholders agree. 
    In this work, we conduct a survey to see how  multiplicity impacts lay stakeholders'
  -- i.e., decision subjects' -- perceptions of ML fairness, and which approaches to address multiplicity they prefer. 
    We investigate how these perceptions are modulated by task characteristics (e.g., stakes and uncertainty). 
    Survey respondents think that multiplicity threatens the fairness of model outcomes, but not the appropriateness of using the model, even though existing work suggests the opposite.  
    Participants are strongly against resolving multiplicity by using a single model (effectively ignoring multiplicity) or by randomizing the outcomes.
    Our results indicate that model developers should be intentional about dealing with multiplicity in order to maintain fairness.
\end{abstract}



\maketitle

\section{Introduction}
Machine learning (ML) is ubiquitous in high-stakes decision-making settings, e.g., social services~\cite{chouldechova2018case}, criminal justice~\cite{compas}, healthcare~\cite{obermeyer2019dissecting}, and hiring~\cite{li2021algorithmic}. 
Due to the life-altering impacts of these decisions (for instance, whether someone can get hired at a job), it is important that the model making the decision is as accurate, fair, and accountable as possible.
The conventional machine learning wisdom is that there exists a single best-performing model and that achieving other desiderata like fairness or robustness requires an accuracy tradeoff~\cite{menon2018cost}. 
But the literature -- both recent~\cite{damourUnderspecification,marxPredictive} and foundational~\cite{breiman2001statistical} -- challenges that paradigm: there are typically many models that achieve similar predictive accuracy by different means (i.e., by making mistakes on different parts of the input distribution). 
This ambiguity leads to \emph{predictive multiplicity}, which is when two equally good models assign different predictions to a given sample, as illustrated by \cref{fig:intro_graphic}. 
In socially-salient prediction settings, the set of (approximately) equally good models is often very large~\cite{marxPredictive,rudin2024position,xin2022exploring}, and a large fraction of samples receive conflicting outcomes under this multiplicity~\cite{marxPredictive}.

Work in computer science and philosophy suggests that multiplicity poses a fairness risk~\cite{black2022model,cooperArbitrariness,creel2022leviathan}. 
Fairness has multiple components, including distributional fairness, which considers whether the \emph{outcomes} are fair, and procedural fairness, which considers whether the decision was made \emph{in a fair way}, regardless of outcomes~\cite{colquitt2001dimensionality}. 
The existing work points out that multiplicity can violate procedural fairness, because choosing one model over the alternatives results in \emph{arbitrary} decisions~\cite{black2022model,cooperArbitrariness,creel2022leviathan}.
This fairness violation is especially salient when models operate in high-stakes settings where the justifiability of a negative decision is important.
For instance, suppose a company uses an ML model $m_1$ to determine which job applicants to interview, but that there is a different model $m_2$ that would do equally as well at selecting promising candidates (e.g., $m_1$ and $m_2$ have similar accuracy on the available validation data).
A candidate $\x$ who is not interviewed according to $m_1$, but who would have been interviewed if $m_2$ were used, may feel that procedural fairness is violated if the company did not even consider using $m_2$. 
An additional concern is that when models are deployed at scale, multiplicity can result in systematic arbitrariness that locks people out of opportunities~\cite{creel2022leviathan}.
For instance, if all companies contract with the same software vendor and each uses model $m_1$ to determine who to interview, $\x$ may get no job interviews even though the existence of $m_2$ suggests they may be qualified.

While the existing literature makes a strong argument for the fairness risks of multiplicity, no research has interrogated whether lay stakeholders believe that multiplicity impacts the fairness of machine learning models. 
Likewise, no research has asked how stakeholders think decisions should be made when a machine learning system contains multiplicity. 
More precisely, consider the same job application scenario where two models $m_1$ and $m_2$ have similar performance, yet disagree on whether to interview a job candidate $\x$. 
The standard machine learning approach is to choose a single well-performing model, for instance, selecting $m_1$ without even realizing that $m_2$ exists.
Philosophers suggest using randomization -- e.g., flipping a coin to determine whether to interview $\x$ -- as the fairest way to deal with multiplicity~\cite{creel2022leviathan,jain2024position}. 
Computer scientists suggest switching to more sophisticated model architectures to reduce multiplicity (e.g., ensemble methods as a way to reduce variance)~\cite{black2022selective,cooperArbitrariness,roth2022reconciling}. 
Lay stakeholders will not necessarily like those approaches, as research shows that different types of stakeholders (like decision makers and decision subjects) can have different perceptions of what matters for fairness in ML~\cite{lee2021who,scott2021exploring,zhang2023Deliberating}.
Critical scholars argue that considering stakeholder preferences is important~\cite{bondi2021envisioning,selbst2019fairness} and this belief underlies frameworks like participatory ML~\cite{feffer2023preference} and value-sensitive design~\cite{umbrello2021mapping}.
Furthermore, companies deploying ML may care about public opinions of fairness to maintain high levels of trust in their product~\cite{woodruff2018qualitative}.

We address the gaps in understanding stakeholders' views on multiplicity by conducting a survey of crowdworkers (i.e., laypeople who are decision subjects as ML is increasingly deployed in everyday settings). Our survey aims to determine (1) how the presence of multiplicity impacts the stakeholders' assessment of ML models' fairness and (2) which multiplicity resolution approaches they prefer.
We consider the following research questions:

\begin{figure*}
    \centering
    \includegraphics[width=14cm]{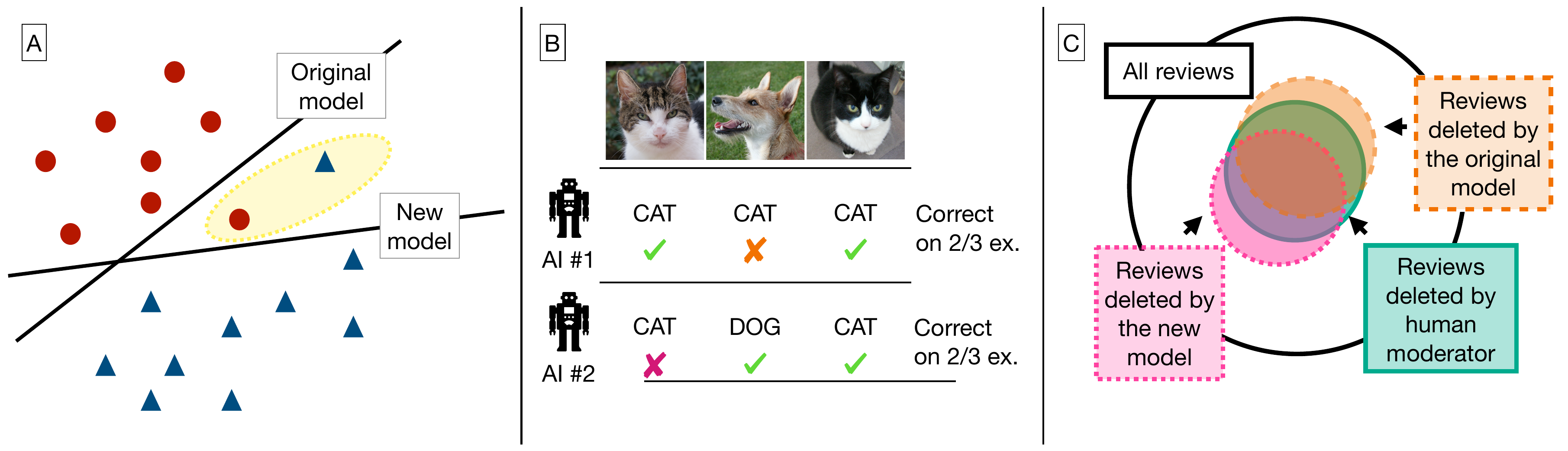}
    \caption{{\textbf{A}: Example dataset with two classes (red circle and blue triangle) that, when limited to linear models, exhibits predictive multiplicity because no model achieves perfect accuracy while multiple models are correct on all but one prediction. The predictions in the yellow shaded region get a different prediction depending on whether we choose the ``original'' or ``new'' model. \textbf{B}: Example of model multiplicity based on model predictions. \textbf{C}: For the task of detecting fraudulent reviews, shows the overlap between reviews flagged as fraudulent by humans and two models. The two models both have significant overlap with human predictions, but less overlap with each other. }  }
    \Description{The image contains 3 sub-figures, labeled A, B, and C. Image A has a cluster of data points (red circles and blue triangles) that are not perfectly linearly separable. The image shows two linear classifiers that each misclassify 1 point (but a different point for each). The points that are misclassified by a model are circled in yellow. 
    Image B is formatted as a table where headers are pictures of animals (Column 1 Cat, Column 2 Dog, Column 3 cat). Each row contains the prediction that an AI system makes on each image (predicting whether it is a cat or dog). The first row (AI 1) is correct on examples 1 and 3, while the second row (AI 2) is correct on examples 2 and 3. Text in the image indicates that both AIs are correct on 2 out of 3 examples.
   Image C is a venn diagram. There is a large circle labeled All Reviews. Within this circle, there are 3 smaller circles arranged along an invisible diagonal line. The middle circle (green) is labeled Reviews Deleted by Human Moderator. The other circles are labeled Reviews deleted by the original model (orange) and Reviews deleted by the new model (pink). The orange and pink circles overlap about 50 percent of their area, while each overlaps with the green circle by about 75 percent of their area.
    }
    \label{fig:intro_graphic}
\end{figure*}

\begin{itemize}
    \item \textbf{RQ1:} Does being alerted to multiplicity in a machine learning system affect participants' perceptions of fairness of the system?
    \item \textbf{RQ2:} What techniques to resolve multiplicity do participants view as most appropriate and fair?
    \item \textbf{RQ3:} How do task characteristics (stakes, uncertainty, and punishment or reward framing) impact \textbf{RQ1} and \textbf{RQ2}? 
\end{itemize}

To address these questions, we want to evaluate fairness perceptions for a variety of tasks. However, some of the factors we care about for \textbf{RQ3} (namely, task stakes and uncertainty) may be subjective, so we did not want to rely solely on our own classification of tasks. Instead, we conducted a preliminary study with $n=199$ participants, which, to the best of our knowledge, is the first study to get lay stakeholders' opinions on both tasks' perceived stakes and uncertainty (Castelo et al.~\cite{castelo2019task} collect opinions on task uncertainty, but not stakes). 
Based on the results of this survey and our own assessment of task framing (reward vs. punishment), we selected 8 diverse tasks to use in the main study. 

Our main study begins with an educational component describing model multiplicity. After the participant passes a comprehension check, we assign them to a scenario and describe a machine learning model at a high level before asking participants for an initial fairness assessment, consisting of a series of Likert scale questions. Then, we describe how model multiplicity impacts the situation and ask for updated answers to the same Likert questions (\textbf{RQ1}). Finally, we use an open-ended question and choice-based analysis to learn what multiplicity resolution techniques participants view as the most fair (\textbf{RQ2}). 

Our results indicate that stakeholders do not perceive multiplicity as an overall fairness risk, counter to what the literature suggests.
However, this result is nuanced as participants are more likely to express that the model outcomes might be unfair, rather than that the model may be inappropriate to use at all.
We also find that participants have clear preferences for how multiplicity is resolved, namely by deferring to a human expert for situations where multiplicity impacts the predictions (\textbf{RQ2}). 
These preferences vary depending on the specific task, with task stakes and task framing significantly impacting resolution preferences (\textbf{RQ3}). 
Despite philosophical arguments for randomization as a multiplicity solution~\cite{creel2022leviathan,jain2024position}, this was the least preferred option in our experiment.

These results have two important implications: first, \emph{the way that multiplicity is discussed in the existing literature does not align with the lay stakeholders' perceptions}. 
 Instead, we hypothesize that other concerns about machine learning (such as recourse and explainability) may dominate participants' fairness assessments.
Second, even though overall fairness perceptions do not shift significantly after learning about multiplicity, the strong sentiments that participants have regarding multiplicity resolution suggest that \emph{ML developers must be deliberate in determining how models behave in the presence of multiplicity}. In particular, simply choosing a single well-performing model and not exploring the impacts of multiplicity is viewed as unfair.

\section{Background of Fairness and Multiplicity}\label{sec:related}
We overview fairness in the context of machine learning (\cref{sec:rel_fairness}), define multiplicity and its associated fairness risks (\cref{sec:mult}), and describe various ways to resolve multiplicity (\cref{sec:mult_solutions}).

\subsection{Fairness in machine learning}\label{sec:rel_fairness}
Colquitt and Rodell propose several components of fairness including distributional, procedural, and informational fairness~\cite{colquitt2015measuring}.
Distributional fairness asks whether a model's outcomes are fair, irrespective of how the decisions were made.
For instance, group fairness metrics such as equalized odds~\cite{dwork2012fairness,hardt2016equality} ask whether different demographic groups are treated equally in aggregate.
Procedural fairness, by contrast, asks whether a decision was made in a fair way, regardless of what the final outcome is. 
Procedural fairness has many facets, including counterfactual fairness~\cite{kusner2017counterfactual} (how a test sample needs to change to get a different prediction) and the choice of features that are included in a model. 
For example, in a task such as hiring, procedural fairness may be violated if race is a decisive factor (i.e., changing an applicant's race changes whether they are hired) or even if race is included as a model input at all.
Informational fairness requires that details about the decision making process be available, and has been explored in ML through the use of explanations, e.g. ~\cite{binns2018s,dodge2019explaining,schoeffer2022there,yurrita2023disentangling}.
In this paper, we consider perceptions of procedural fairness.

The fairness of a ML model can be measured based on its adherence to some fixed standard (e.g., a group fairness metric), or in a human-centered way through perceived fairness, as in this study. Perceived fairness is a common way to evaluate ML systems~\cite{starke2022fairness}, perhaps because settling on a formal fairness definition is challenging. Research has established that perceived fairness of a model correlates with trust of that model~\cite{shin2020beyond,shin2019role,woodruff2018qualitative}, and suggests that low perceived fairness corresponds to negative consequences (e.g., reputational harm) for the entity deploying ML~\cite{starke2022fairness}. 

\subsection{Multiplicity in machine learning and fairness concerns}\label{sec:mult} 

Predictive multiplicity~\cite{marxPredictive}, which we simply call \emph{multiplicity} for brevity, occurs when there are multiple ML models that are equally well-suited for a task, yet give different individual-level predictions. 
Formally, suppose we have a dataset $D$, a hypothesis class $\mathcal{H}$ (e.g., all possible decision trees with depth $\leq 3$) and some loss function $\mathcal{L}$ (e.g., the 0-1 loss). 
A standard machine learning pipeline aims to find $h^*\in\mathcal{H}$  that minimizes $\mathcal{L}_{h}(D)$ (e.g., the total number of misclassifications on $D$). 
This solution is typically found via an approximate optimization process such as gradient descent, and yields a model $h_1$ that hopefully has close-to-optimal loss, i.e., has $\mathcal{L}_{h_1}(D)\approx \mathcal{L}_{h^*}(D)$. 
But sometimes, the optimization problem is underconstrained and there is another model $h_2$
such that $\mathcal{L}_{h_2}(D)\approx \mathcal{L}_{h_1}(D)$. (Intuitively, there may be multiple global minima.)  
If there is some input $\x\in D$ with $h_1(\x)\neq h_2(\x)$ we say that the models exhibit \emph{(predictive) multiplicity}.

\paragraph{Prevalence of multiplicity} 
Computing the fraction of inputs whose predictions are affected by multiplicity is generally intractable, so
existing approaches quantify multiplicity by limiting their analysis to specific model types, approximating the impacts of multiplicity, and/or using brute force. 
Estimates on the COMPAS dataset, which is used to predict risk of recidivism in defendants, indicate that multiplicity affects 44\% of test inputs when using linear models~\cite{marxPredictive} and 69.4\% of test inputs when using decision trees~\cite{cooperArbitrariness}. 
For probabilistic models that compute risk scores, model outputs vary by $>10\%$ on 82-100\% of samples across three tabular datasets~\cite{watson2023predictive}. 
These levels of multiplicity indicate that many predictions are a function of the specific model that is chosen, rather than purely a function of the data.

\paragraph{Fairness risks of multiplicity}
Fairness risks of multiplicity include harms to an individual, society or a group of individuals, and the entity who is deploying the model.  
Fairness risks to individuals include a lack of justifiability for a negative decision~\cite{black2022model} and systemic loss of opportunity when models containing multiplicity are deployed at scale~\cite{creel2022leviathan}. 

\begin{example}[Lack of justifiability]
   Black et al~\cite{black2022model} point out that US credit laws like the Equal Credit Opportunity Act aim to reduce arbitrariness and increase equity in credit decisions, such as by requiring lenders to provide justification for denied loan requests. However, if another multiple equally-accurate model approves a loan request, it may be more challenging to justify why the  bank needed to use the model that denies the loan.
\end{example}

\begin{example}[Systemic loss of opportunity]
    Creel and Hellman~\cite{creel2022leviathan} discuss how individual employers may have idiosyncratic hiring criteria, but that the harm from this to job applicants is minimal, as they can apply to many jobs. Since each employer has their own potentially arbitrary criteria (in other words, multiplicity exists in the process of choosing who to hire), the job applicant will eventually be hired by someone. However, if every company uses the same hiring system, then certain job applicants will not be able to be hired anywhere, causing them significant harm. This setup can happen, for instance, if every company contracts with the same third-party software to screen resumes. 
\end{example}
Society or groups of individuals 
may be adversely impacted by multiplicity if a sub-optimal model, with regards to fairness, is selected from the set of equivalently-accurate models.
The entity using the model is also at risk, first for potential lawsuits or reputational harm if their decisions are not justifiable~\cite{black2022model}. Second, if a single model is deployed at scale when multiplicity exists, this standardization can provide a worse set of outcomes.
\begin{example}[Worse outcomes for the decision-maker]
Kleinberg et al.~\cite{kleinberg2021algorithmic} show that companies who all use the same recruiting software may end up with lower-quality new hires than if each company used its own model, an outcome that is harmful for the company itself and not just potential job applicants.
\end{example}
However, no work that we are aware of has investigated the fairness risks of multiplicity \emph{from the perspective of stakeholders}.
Previous work shows that formal fairness definitions often do not line up with stakeholders' fairness perceptions~\cite{srivastava2019mathematical}, so, it seems likely that experts and lay stakeholders may have distinct insights about whether multiplicity is (un)fair.

\subsection{``Solving'' multiplicity}\label{sec:mult_solutions}

A natural question, after finding evidence of multiplicity, is how to go about making decisions. 
In this paper, we consider the fairness of six different multiplicity-resolution techniques (see \cref{tab:part2_options}). 
The standard approach in machine learning is to \texttt{ignore}\footnote{Here, and throughout the rest of the paper, we use this font to denote the names of the multiplicity-resolution techniques.} multiplicity; that is, to find a single well-performing model and use it, without considering whether multiplicity exists. Of work that focuses on multiplicity, three schools of thought exist on how to resolve it: reducing multiplicity through algorithmic interventions, standardizing the procedural response to multiplicity, and using multiplicity as an opportunity to optimize for second-order criteria like fairness or robustness.

\paragraph{Reducing multiplicity}
Refining the modeling technique can reduce multiplicity, such as by using cascading classifiers~\cite{roth2022reconciling}, bootstrapping~\cite{cooperArbitrariness}, or selective ensembles that abstain if the confidence is not high enough~\cite{black2022selective}. 
Similar ensembling approaches have long been used to improve accuracy in ML applications~\cite{sagi2018ensemble}.
The difference when using ensembling to combat multiplicity is that multiplicity is often discussed in the context of social prediction tasks where an objective ground truth is elusive or even does not exist.
Thus, using ensembling to get rid of multiplicity often does not yield significantly higher accuracy than using a non-ensembled model. It does, however, mask that multiplicity exists, making it harder to challenge the validity of an algorithm.
We refer to ensembling and other approaches that aim to reduce multiplicity through modeling choices as \texttt{fancy}, i.e., training a fancier (more complex) model.

\begin{wraptable}{r}{0.49\textwidth}
\caption{The multiplicity resolution techniques that we consider. We explain to participants that the proposed method is only used if the two AI systems disagree. We describe the techniques in more detail tailored to the specific task (e.g., ``Let a doctor decide'' for \texttt{human} in a medical setting); full text is in the appendix.}\label{tab:part2_options}
\begin{tabular}{ll}\\\toprule  
\textbf{Method name} & \textbf{When the two models disagree...} \\\midrule
\texttt{default-bad} & Default to the undesirable outcome \\  
\texttt{default-good}  & Default to the desirable outcome \\  
\texttt{fancy} & Train a more complex model \\  
& and use its output \\
\texttt{human} & Let a human expert decide \\
\texttt{ignore} & Use the original model's decision \\
\texttt{random} & Choose the outcome randomly \\\bottomrule
\end{tabular}
\end{wraptable} 

\paragraph{Dealing with multiplicity procedurally}
ML researchers have suggested abstaining, and, presumably, deferring to a human expert (denoted in this paper as \texttt{human}), when a decision is subject to multiplicity~\cite{black2022selective}. 
However, the design of ``human-in-the-loop systems'' is not straightforward~\cite{wu2022survey}, and may interact with other criteria we are trying to measure, e.g., algorithm aversion~\cite{fogliato2022goes}.
The idea of abstaining and deferring to a human decision-maker approach parallels a long line of work in robust machine learning that similarly suggests abstaining on decisions when robustness cannot be verified~\cite{bartlett2008classification,cohen2019certified,herbei2006classification,laidlaw2019playing}.\footnote{Note that non-robustness constitutes a form of multiplicity, as it implies there is another reasonable model that outputs an alternate prediction for the test sample. Within the ML literature, robustness is typically discussed for deep networks and/or text and image data, while multiplicity research largely focuses on social prediction tasks using tabular data.} 
However, these works rarely present a plan for what abstaining looks like in practice: often, abstaining on a decision results in not taking action, thereby defaulting to the negative label (for instance, abstaining on a loan application is effectively denying it). 
We consider abstaining and deferring to a default option in our survey. 
However, attitudes may vary depending on whether the outcome is desirable or not, so we use \texttt{default-good} and \texttt{default-bad} to represent the outcomes that are good and bad, respectively, from the perspective of the decision subject.

An alternate procedural approach is \texttt{random}, i.e., randomizing over the set of equally-good models~\cite{creel2022leviathan, jain2024position}. 
ML in social prediction tasks often serves to distribute a limited resource, e.g., admissions to a college with a fixed enrollment limit. Lotteries are typically viewed as fair system for dividing limited resources~\cite{broome1990fairness,sher1980makes}. However, there are technical challenges in implementing randomization effectively~\cite{jain2024position}. 

\paragraph{Optimizing for second-order criteria} 
Traditional ML practice assumes that there is a trade-off between model accuracy and fairness, as defined by group-level metrics such as equalized odds or statistical parity~\cite{kamiran2012data,menon2018cost}.
Recent research to explore the whole set of equally-good models challenges this paradigm, showing that accuracy and other criteria are not in tension~\cite{rudin2024position,xin2022exploring}.
But even if we can use multiplicity to find fairer (or more robust) models, there are still multiple mutually incompatible fairness (or robustness) definitions.  
We do not include optimizing for fairness as one of the possible multiplicity resolution techniques because the task of choosing the fairest model is typically not well-defined~\cite{chouldechova2017fair,kleinberg2017inherent}.

\section{Related Work and Hypotheses}\label{sec:hypotheses}

Fairness perceptions of machine learning algorithms have been studied extensively, as evidenced by multiple survey papers on the topic~\cite{narayanan2024fairness,starke2022fairness}. Here, we will focus on dimensions most related to our research questions.

Research in this area investigates distributional fairness (e.g., matching lay stakeholder's fairness notions to formal definitions~\cite{saxena2019how,srivastava2019mathematical}), procedural fairness (e.g., considering what features to include~\cite{grgichlaca2018human}), and informational fairness (e.g., the impact of explanations~\cite{schoeffer2022there,yurrita2023disentangling}).
In this work, we focus on perceptions of procedural fairness, presenting multiplicity as a hurdle that must be dealt with procedurally. 
The fairness perceptions of ML models are often considered relative to a baseline of human decision making, and 
can either increase from algorithmic appreciation~\cite{logg2019algorithm}, or decrease from algorithmic aversion~\cite{dietvorst2015algorithm}. 
Existing work establishes that fairness perceptions of ML models vary based on task characteristics, including stakes~\cite{hannan2021gets,saragih2022effect,yurrita2023disentangling},  uncertainty~\cite{salimzadeh2024dealing}, complexity~\cite{salimzadeh2023missing}, and framing (reward vs. punishment)~\cite{hannan2021gets}.

In the following subsections, we describe our hypotheses about how multiplicity impacts perceived fairness (\hOneMain-\hOneUnc) and which multiplicity resolution techniques stakeholders prefer (\hTwoMain-\hTwoUnc). See \cref{tab:hyps} for a summary of all hypotheses. 

\subsection{Algorithm aversion}
Algorithm aversion is when people prefer human decision-makers, even when the available AI system generally outperforms humans~\cite{dietvorst2015algorithm}. 
It may be triggered when people see a model make a mistake, even if they would have been forgiving of a human making the same mistake~\cite{dietvorst2015algorithm}, or when the model's performance is worse than stated~\cite{yin2019understanding}. 
In settings where a correct answer can take many forms (e.g., generative AI), seeing multiple conflicting output lowers trust in a model~\cite{lee2024one}.
Algorithm aversion also tends to be higher for subjective tasks, a characterization that is malleable depending on how the task is described~\cite{castelo2019task}.
We expect that witnessing model multiplicity will increase algorithm aversion for three reasons: first, the existence of two good (but different) models means that the models make mistakes, similar to what was observed for conflicting generative AI responses~\cite{lee2024one}. 
Second, the existence of multiple good models indicates that the decisions these models make are subjective. In particular, multiplicity means that choosing a model is not a purely objective task (i.e., optimizing accuracy), but rather a subjective choice of choosing between multiple models with near-equivalent accuracy.
Finally, existing work assumes (without consulting stakeholders) that multiplicity threatens fairness (\cref{sec:mult}).

\hOneMain: Stakeholders will view machine learning systems as less fair once they are made aware of multiplicity.

Between algorithm aversion and prior work showing preference for human over AI arbitrators~\cite{sela2018can}, we think participants will prefer \texttt{human} over other resolution techniques 
(recall that \texttt{human} and the other approaches are defined in \cref{tab:part2_options}). 
We think that -- counter to what philosophical work suggests -- \texttt{random} will be viewed as the least fair approach. Explanations are important to fairness~\cite{schoeffer2022there,yurrita2023disentangling}, so we think the 
sense of unpredictability from randomized decisions dominate stakeholders' reactions to \texttt{random}.

\hTwoMain: Stakeholders will have significant preferences between different multiplicity solutions. In particular, they will view \texttt{human} as the fairest multiplicity solution and \texttt{random} as the least fair solution.  

\subsection{Task stakes} 
Competing with algorithm aversion is \emph{algorithm appreciation}, when people over-rely on algorithmic decisions~\cite{logg2019algorithm}. 
There is evidence that algorithm appreciation occurs in high-stakes tasks~\cite{araujo2020ai,saragih2022effect}, and stakeholders rate accuracy as more important in high-stakes settings~\cite{srivastava2019mathematical}. 
Thus, we think that \texttt{fancy} (i.e., building a more complex model) may be preferred for high-stakes tasks because it may be perceived as being more accurate.
However, the importance of accuracy may also trigger algorithm aversion and increase preference for \texttt{human}.
Fairness is more important to stakeholders in high-stakes settings than low-stake ones~\cite{jakesch2022different}. 
We suspect that higher-stakes tasks are more emotionally charged. People view emotional decision making as a strength of humans~\cite{helberger2020fairest}, and can have negative emotional responses to AI decision-making~\cite{lee2018understanding}. 

\hOneStakes: The presence of multiplicity will have a larger impact on fairness perceptions in high-stakes settings.

\hTwoStakes: Participants' preferred multiplicity resolution technique will be moderated by task stakes. In particular, high-stakes tasks will correspond to stronger preferences for \texttt{human} and \texttt{fancy}, and low-stakes tasks will correspond to stronger preferences for \texttt{ignore} and \texttt{random}.

\subsection{Task framing} 
Harm that is caused by action, rather than omission, causes harsher fairness judgments~\cite{cushman2006role}. 
Tasks that are punitive commit active harm in the event of a mistake, whereas reward-based tasks cause harm when they make an omission. 
Thus, we think that punitive tasks will have greater decreases in fairness perceptions in the presence of multiplicity. 
Indeed, others have found that when people adopt critical mindsets,
they are more likely to find issues with a model~\cite{hannan2021gets}.  
Algorithms are seen as more rational than humans~\cite{logg2019algorithm}, and we suspect that rationality will be more important in punitive settings where mistakes are judged more harshly. 
Conversely, there are positive connotations towards human accuracy~\cite{lee2018understanding}, so being primed to think about reward settings may influence participants to select \texttt{human}.

\hOneFraming: The presence of multiplicity will have a larger impact on fairness perceptions in punitive settings. 

\hTwoFraming: Preferences about how multiplicity is resolved will be moderated by task framing. In particular, for punitive tasks, \texttt{fancy} will be more preferred and for reward-based tasks, \texttt{human} and \texttt{default-good} will be more preferred. 

\subsection{Task uncertainty} 
High-uncertainty tasks typically do not have a single objectively-right answer, such as predicting future human behavior (e.g., job performance) or matters of taste (e.g., movie recommendation). Low-uncertainty tasks typically have an objectively-correct answer (e.g., detecting spam emails). 
Algorithm appreciation is known to be stronger in objective (i.e., low-uncertainty) tasks~\cite{bankins2022ai,logg2019algorithm}, aligning with beliefs that algorithms are not good at subjective (i.e., high-uncertainty) tasks~\cite{castelo2019task}.  There is more algorithm aversion and preference for human decision makers in high-uncertainty settings~\cite{dietvorst2020people}. 
When a task has lower uncertainty, it may be possible to have overall higher accuracy, which increases algorithm appreciation~\cite{saragih2022effect}.
Intuitively, people may find it easier to rationalize that there are multiple good solutions to a high-uncertainty task, resulting in lower impact to the perceived fairness. 

\hOneUnc: The presence of multiplicity will have a larger impact on fairness perceptions for low-uncertainty tasks.

\hTwoUnc: Preferences of how to resolve multiplicity will be moderated by task uncertainty. In particular, for low-uncertainty tasks stakeholders will see \texttt{fancy} as more fair, while for high-uncertainty tasks, they will see \texttt{random} as more fair.

\section{Method}\label{sec:method}

\begin{figure*}
    \centering
    \includegraphics[width=14cm]{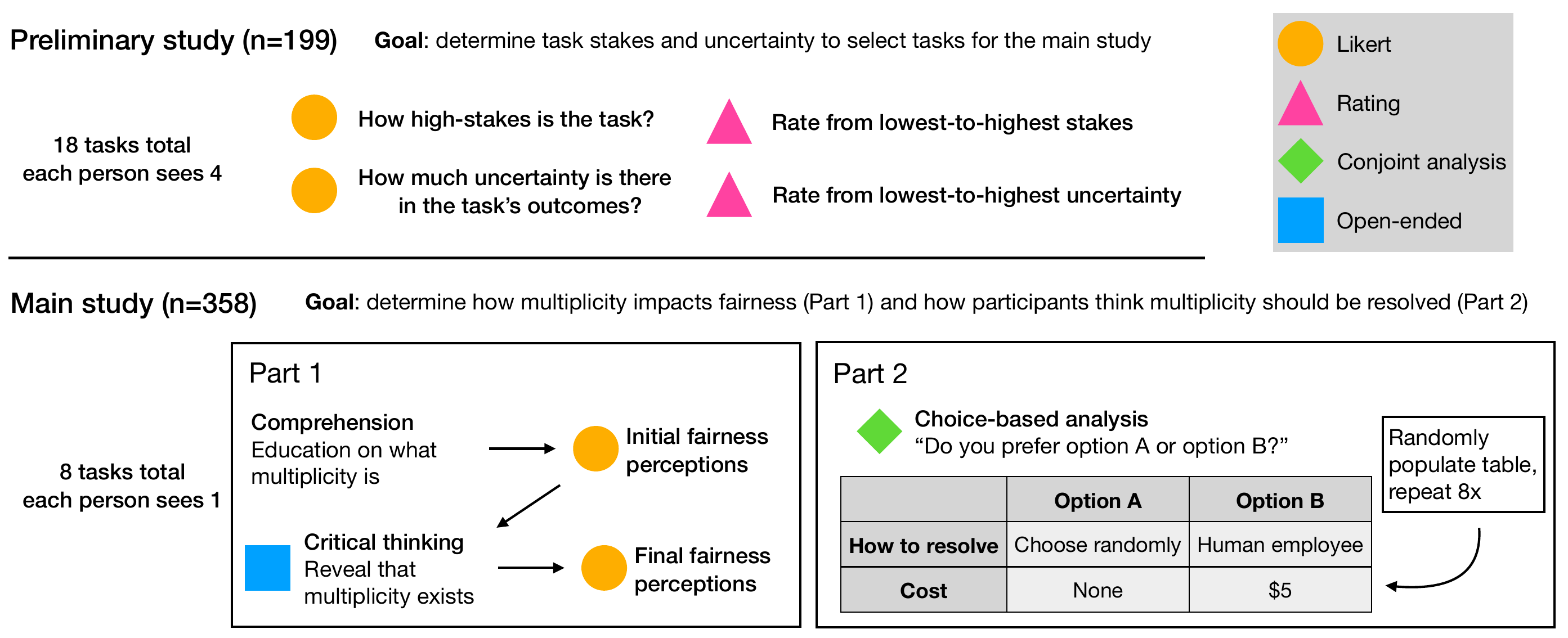}
    \caption{ Overview of our survey design, including the preliminary study (top row) and main study (bottom row).}
    \Description{Visual overview of our study design. There are two rows. First row text is Preliminary study (n=199). Goal: determine task stakes and uncertainty to select tasks for the main study. There are 18 tasks total and each person sees 4. We ask How high-stakes is the task (Likert), How much uncertainty is there in the task outcomes (Likert), and we ask the participants to rank the 4 tasks they see on stakes and on uncertainty.
    The second row says Main Study (n=358). Goal: determine how multiplicity impacts fairness (Part1) and how participants think multiplicity should be resolved (Part 2). There are 8 tasks total and each person sees 1. The remainder of this row is subdividing into Part 1 and Part 2. In part 1, the flow is Comprehension (education on what multiplicity is) to Initial fairness perceptions (Likert) to Critical thinking (reveal that multiplicity exists, open-ended response), to Final fairness perceptions (Likert).
    In Part 2, it says Choice Based Analysis (we ask do you prefer option A or Option B). We randomly populate a table with two rows and two columns. The first column is labelled option A and the second column is labelled option B. The first row is How to resolve multiplicity and the second row is Cost. The example shown in the image has Option A= Choose randomly and None and Option B = Human employee and 5 dollars. 
    }
    \label{fig:overview}
\end{figure*}

\Cref{fig:overview} shows an overview of our study design, which was approved by our institution's IRB board. In this section, we first describe our preliminary study (\cref{sec:method_prelim}), which informed the task selection for our main study. Then, we describe our pre-registered study\footnote{\url{https://osf.io/fv3bt}} design (\cref{sec:study_design}) and data analysis plan (\cref{sec:method_stats}). 

\subsection{Task selection (Preliminary study)}\label{sec:method_prelim}
Our main study asks participants about tasks that differ on three between-subjects axes: stakes, uncertainty, and framing. Framing (i.e., whether a positive model output is a reward or a punishment) is objective; however, individuals may differ on how high-stakes or uncertain they think a task is.
We conduct a preliminary study ($n=199$) to solicit perceptions of task stakes and uncertainty. 
Our goal is to select one task for each of the eight categories [punishment, reward] $\times$ [low uncertainty, high uncertainty] $\times$ [low stakes, high stakes]. 

\paragraph{Study design}
We collect 18 scenarios through brainstorming and borrowing from previous studies~\cite{binns2018s,castelo2019task,kasinidou2021agree}.
We randomly assign each participant to see 4 scenarios to limit participant fatigue when completing the study. 
For each scenario, participants answer a series of 5-point Likert-scale questions relating to task clarity (2 questions), task stakes (3 questions), and task uncertainty (2 questions). 
Then, participants rank the four tasks according to stakes and uncertainty levels. 
We use Prolific to find respondents (US-based adults with a 97\%+ approval rate). We compensate participants at Prolific's recommended rate (\$12/hour) based on the estimated survey completion time. 

\paragraph{Data analysis} 
Our survey yields four pieces of data: stakes rating, uncertainty rating, stakes ranking, and uncertainty ranking.
We convert all Likert ratings to a scale from -2 (strongly disagree) to 2 (strongly agree).
For the rating data, we average the three questions relating to task stakes into a single stakes rating quantity. We elect to use one of the task uncertainty questions alone, due to better data quality. 
To recover the underlying stakes and uncertainty rankings given the partial ranking information from each participant, we use assume a Plackett-Luce distribution~\cite{luce1977,plackett1975} and use the I-LSR (iterative Luce Spectral Rating) algorithm~\cite{maystre2015fast}. 
The basic assumption behind Plackett-Luce is that choice rankings are transitive (i.e., if A>B and B>C then A>C), and the I-LSR algorithm uncovers the full underlying ranking.
We combine the rating and ranking data into a single metric by scaling each set of data to have mean 0 and variance 1, and then taking their arithmetic mean. 

\paragraph{Preliminary study results}
We obtain responses from 201 participants and, after excluding responses that failed more than one attention check, use 199 responses in the analysis.

Task stakes and task uncertainty are not meaningfully correlated ($r$=0.19, $p$=0.42). 
We exclude three tasks that scored poorly on clarity metrics, leaving 15 tasks to choose from. 
There are trade-offs in terms of achieving either a wide range of scores on each axis or relative parity between tasks in different (stakes $\times$ uncertainty) quadrants. That is, we can either select tasks to maximize the range of stakes and uncertainty scores, or we can select tasks with a narrower range of scores but with a better one-to-one comparison between tasks with different framing. 
We decide to maximize the range stakes and uncertainty scores since our analysis plan involves using numeric stakes and uncertainty labels rather than solely binary ``low'' and ``high'' categories. 
The tasks we select are described in \cref{tab:tasks}.

\begin{table*}[t]
\caption{Summary of tasks and their stakes, uncertainty, and framing. Starred columns (*) are based on the results of our preliminary study (\cref{sec:method_prelim}). (The raw stakes and uncertainty scores are binarized based on whether they are above or below the medians across all 18 tasks from that study.)
A full description of tasks and their associated stakes and uncertainty numeric values are available in the appendix. }
\label{tab:tasks}
\begin{tabular}{ll|ccc}\toprule
Domain & Task & Stakes* & Uncertainty* & Framing     \\\midrule
Healthcare & Access to a preventative health care plan & High & High & Reward     \\
Business & Job probation based on employee monitoring & High & High & Punishment \\ 
Healthcare & Tumor detection from medical imaging & High & Low  & Reward     \\
Government & Tax audit & High & Low  & Punishment \\
Travel & Airline seat upgrade & Low  & High & Reward     \\
Travel & Rebooking passengers on an over-full flight & Low  & High & Punishment \\
Business & Customers get beta access to an app with discounts & Low  & Low  & Reward     \\
Business & Removing online reviews that are spam or fake & Low  & Low  & Punishment \\
\bottomrule     
\end{tabular}
\end{table*}

\subsection{Study design}\label{sec:study_design}
The survey flow is depicted in \Cref{fig:overview}. 
The survey begins with education about multiplicity (with a comprehension question) and then has two main sections (``Part 1'' and ``Part 2''). 
Participants are randomly assigned to one task at the beginning of Part 1. 
Then we describe a scenario and solicit the participant's fairness perceptions of the model on a Likert scale consisting of four seven-point questions. 
In Part 2, we present a series of binary choices between different ways to resolve multiplicity and the user selects their preferred option. 
We conclude by collecting demographic information from the participant.
We provide further details on all parts of the survey below.

\paragraph{Details of multiplicity education}
We describe to participants, at a high level, how AI models are built (e.g., ``AI developers [...] choose the model with the best performance''). Then, we reveal that there are sometimes multiple models that can perform equally well, but ``make different mistakes''. We illustrate model multiplicity using a graphic similar to \cref{fig:intro_graphic}B. We tell participants that this phenomenon, model multiplicity, is common. Then, participants have two chances to correctly answer the multiple-choice question ``What does model multiplicity mean in the context of AI?'' The correct answer is written using similar phrasing as in the descriptive text. We include full text of the multiplicity education section in the appendix.

\paragraph{Details of Part 1}
In Part 1, we aim to address \textbf{RQ1} and \textbf{RQ3} by seeing how participants' fairness perceptions of ML systems change when multiplicity is present.
We first describe the scenario including what decision is being made, what factors the AI has access to, and how decisions will be operationalized (e.g., whether users will have recourse).
We also say that the AI is validated and ``agrees with human experts most of the time''. 
This description of accuracy is vague to avoid anchoring participants to specific numbers and because reporting an ``accuracy'' suggests a ground truth that does not exist for all scenarios we consider.

Next, we ask users to respond to four seven-point Likert scale questions to elicit fairness perceptions. 
Two questions focus on the model outcomes, e.g., whether the proposed model is likely to be able to make ``good'' decisions, and two questions focus on whether using the model is appropriate. 
Then, we describe how multiplicity impacts the situation, 
explaining that there is a second model that agrees with human experts equally often as the first model, 
but disagrees less often with the original model. 
We accompany the description with a graphic such as the one shown in \Cref{fig:intro_graphic}C. 
At this point, participants answer two critical-thinking questions:
one is multiple-choice and comprehension based; the other is open-ended and asks how decisions should be made when the two models disagree. 

The next page of the survey asks the participant to respond again to the same set of Likert questions. 
We refer to the first set of responses as the ``pre'' timestamp, and this set of responses as the ``post'' timestamp, because the latter are collected after the participant learns that multiplicity exists for the task.
The participant does not have access to their previous responses at the ``post'' timestamp, because we 
do not want them to be anchored by their original responses (i.e., we want to allow for a change in their frame of reference~\cite{howard1979internal}).

\paragraph{Details of Part 2}
In part 2, we investigate \textbf{RQ2} and \textbf{RQ3} by seeing which multiplicity resolution techniques participants prefer.
We provide a description of 5-6 ways (depending on the task) that multiplicity can be resolved, i.e., how to make decisions when the two models disagree (see \cref{tab:part2_options}). 
For some tasks, \texttt{default-good} is not realistic due to logical constraints, e.g., airline seat upgrades likely cannot be given to every passenger selected by either model because there is a firm limit in the number of available first-class seats. 
In cases like this, we exclude that option. 
We also explain to participants that some options may be more expensive to implement, 
so might require a trade-off such as a monetary cost. 
The cost that participants see for each option is either ``no extra cost'' or a small scenario-specific cost that is meant to be an annoyance, but not a barrier for most people (e.g., airline tickets will cost \$5 extra). 

The participant then sees 8 randomly constructed binary choices between two randomly chosen multiplicity resolution techniques and associated costs, such as the example in \cref{fig:overview}. 

\paragraph{Demographic information}
We collected demographic information from participants including gender, age, education level, and experience with AI. Experience with AI was collected through the question ``Do you interact with artificial intelligence or automated decision-making systems as part of your work or school?'', with the choices ``yes,'' ``no,'' and ``not currently, but in the past''.

\subsection{Data collection}
We collect data using Prolific. Our participation criteria is US-based adults (18+) who have at least a 95\% approval rate. 
We plan to collect data from 360 participants, i.e., 45 per task. We estimated that 35-40 participants per task would be sufficient based on trends in pilot data; we rounded up to 45/task to have additional buffer in case we need to reject some samples. 
We follow Prolific's guidance~\cite{prolific} and exclude survey respondents who fail the initial comprehension check question twice, fail two or more attention check questions, or are statistical outliers in survey completion time (in the fast direction).

\subsection{Statistical analyses}\label{sec:method_stats}

\subsubsection{Part 1 analysis}
\paragraph{Data transformation} We recode the Likert scale data from -3 (strongly disagree) to 3 (strongly agree) and compute the average across the four Likert scale questions.

\paragraph{Testing \hOneMain-\hOneUnc}
To test \hOneMain-\hOneUnc, we fit a linear mixed model (LMM), 
which generalizes a linear model by accounting for both between- and within-subject factors. 
The formula we fit to the LMM predicts the transformed Likert data as a function of time, task stakes, task framing, task uncertainty, respondent ID (to account for individual differences), and the first-order interaction terms of time with each task characteristic.
We use the \texttt{lme4} package~\cite{bates2015Fitting} in R to construct the LMM model 
and then use the \texttt{car} package~\cite{fox2019car} to conduct a type-III ANOVA. 
Using ANOVA procedures on Likert-scale data is controversial because Likert scales are not continuous and may not produce normally distributed data. 
Scholars have argued both for~\cite{blanca2017non,carifio2008resolving,norman2010likert} and against~\cite{jamieson2004likert,kuzon1996seven} applying ANOVA when its assumptions are violated. 
We decided that LMMs fit our needs best, so to minimize concerns about using ANOVA, we included a large population in each group (as required by the central limit theorem for normality of the data sample) and used the average of several Likert items rather than a single item~\cite{carifio2008resolving}. 
The main alternative technique we considered is Aligned-Rank-Transform (ART) ANOVA~\cite{wobbrock2011aligned}, a procedure that transforms ordinal (e.g., Likert-scale) data to fit the ANOVA assumptions.
Ultimately we chose LMMs because ART-ANOVA does not allow for continuous factors. 
However, for readers who are interested, we include two additional analyses in the appendix: ART-ANOVA and a non-parametric test, the Mann-Whitney U-Test~\cite{mann1947test}.\footnote{To clarify, we preregistered the analysis with LMMs. Readers should view the results of the other tests as a partial \emph{multiverse analysis}~\cite{steegen2016increasing}.}

To test \hOneMain, we check whether the ANOVA p-value is significant, and if so, run a post-hoc contrast test to gauge directionality. 
To test \hOneStakes and \hOneUnc, we find the estimated marginal trends for how task stakes (\hOneStakes) and uncertainty (\hOneUnc) vary with time using \texttt{emtrends} in the \texttt{emmeans} package in R~\cite{searle1980population}. Then, we conduct a pairwise contrast test on the result. To test \hOneFraming, we perform a pairwise contrast test on the estimated marginal means of the interaction term of time and task framing, again using \texttt{emmeans}. 

\paragraph{Multiple hypothesis correction} In total, since we perform four post-hoc contrast tests on the Likert scale data, we will use the Holm-Bonferroni correction~\cite{holm1979simple} with $\alpha=0.05$ and $m=4$ to evaluate the hypotheses.

\subsubsection{Part 2 analysis}

\paragraph{Conjoint analysis} 
We test \hTwoMain-\hTwoUnc using choice-based conjoint analysis~\cite{hainmueller2014causal}. Choice-based analysis is a technique that solicits user preferences over two options with various factors set to different attribute levels. The analysis generates the marginal means of each attribute level. Our analysis uses the \texttt{cregg} package in R~\cite{cregg}.  For \hTwoMain, we look directly at the marginal mean effect of each of the multiplicity resolution techniques and costs. For each condition, the marginal mean procedure outputs a point estimate and a confidence interval of how preferred an option is relative to random chance. For instance, an output of 0.8 indicates that participants presented with that option chose it 80\% of the time. For \hTwoStakes, \hTwoFraming, and \hTwoUnc, we compute the marginal means stratified by task characteristics. To the best of our knowledge, there is no tool to compute marginal means for a continuous factor, so we use the binary encodings (low vs. high) for task stakes and task uncertainty. 

\paragraph{Testing \hTwoMain}
We will consider \hTwoMain to be weakly supported if the confidence interval for \texttt{human} is strictly greater than 0.5 and the confidence interval for \texttt{random} is strictly less than 0.5. 
Then, we will test whether the estimate for \texttt{human} is strictly higher than the estimates for the other methods. To do this, we will perform five two-sided t-tests\footnote{The choice-based analysis data may not be normally distributed; however, we have over 800 observations for each resolution technique so the central limit theorem makes the use of this test acceptable.} (one against each other multiplicity resolution technique).
Then, we will perform four t-tests (\texttt{random} vs. each other multiplicity resolution technique other than \texttt{human}) to see whether \texttt{random} is strictly less preferred than each other technique. 

\paragraph{Testing \hTwoStakes-\hTwoUnc}
We compute the marginal means model for each task characteristic. 
We then perform post-hoc t-tests on the estimates. 
For instance, to test \hTwoStakes, we first check whether the estimate for \texttt{human} for low-stake tasks is different from its estimate for high-stake tasks. 
In total, we perform four post-hoc tests for \hTwoStakes, three for \hTwoFraming, and two for \hTwoUnc.

\paragraph{Multiple hypothesis correction}
Since there are six multiplicity resolution techniques, we use the Bonferroni\footnote{Note, we do not use Holm-Bonferroni here because it is not supported by \texttt{cregg}.} correction to get a p-value of $0.0083$ and use this to compute the confidence intervals for each marginal means estimate. 
We perform a total of 18 post-hoc t-tests on the data, so will evaluate their significance using the Holm-Bonferroni correction with $\alpha=0.05$ and $m=18$.

\section{Data Analysis}\label{sec:results}

\begin{wraptable}{r}{0.535\textwidth}
\caption{Summary of the hypotheses described in \cref{sec:hypotheses}. The final column indicates whether we found statistically significant support for the hypothesis (\cref{sec:res_hypo})} 
\label{tab:hyps}
\begin{tabular}{c|ll}\toprule
 & Description & Sig? \\\midrule
 \hOneMain & Multiplicity reduces fairness perceptions & No \\
 \hOneStakes & \hOneMain is modulated by task stakes & No \\
 \hOneFraming & \hOneMain is modulated by task framing & No \\ 
 \hOneUnc & \hOneMain is modulated by task uncertainty & No \\
 \hTwoMain & Participants care how multiplicity is resolved & Yes \\
 \hTwoStakes & \hTwoMain is modulated by task stakes & Yes \\ 
 \hTwoFraming & \hTwoMain is modulated by task framing & Yes \\
 \hTwoUnc & \hTwoMain is modulated by task uncertainty & No \\\bottomrule
    \end{tabular}
\end{wraptable}

\subsection{Summary of responses}\label{sec:res_ummary}
We receive 367 survey responses and after excluding ones that fail the initial comprehension check twice or fail multiple attention checks, we use 358 in our analysis. 58.7\% of respondents are women, 2.0\% are non-binary, and 39.4\% are men. 
The average respondent age is 36.3 (range: 18-78, stdev: 12.3). 
For the ``pre'' timestamp, Cronbach's Alpha between the four Likert items is 0.88 and for the ``post'' timestamp it is 0.94.

\subsection{Evaluation of hypotheses}\label{sec:res_hypo}

We find no evidence for \hOneMain ($\chi^2(1)=0.146, p=0.703)$. \hOneStakes ($\chi^2(1) = 1.262, p=0.261$), 
\hOneFraming ($\chi^2(1) =0.308, p=0.579)$, 
or \hOneUnc ($\chi^2(1) = 1.405, p=0.236$).

\begin{wrapfigure}{r}{0.5\textwidth}
    \begin{center}
            \includegraphics[width=0.48\textwidth]{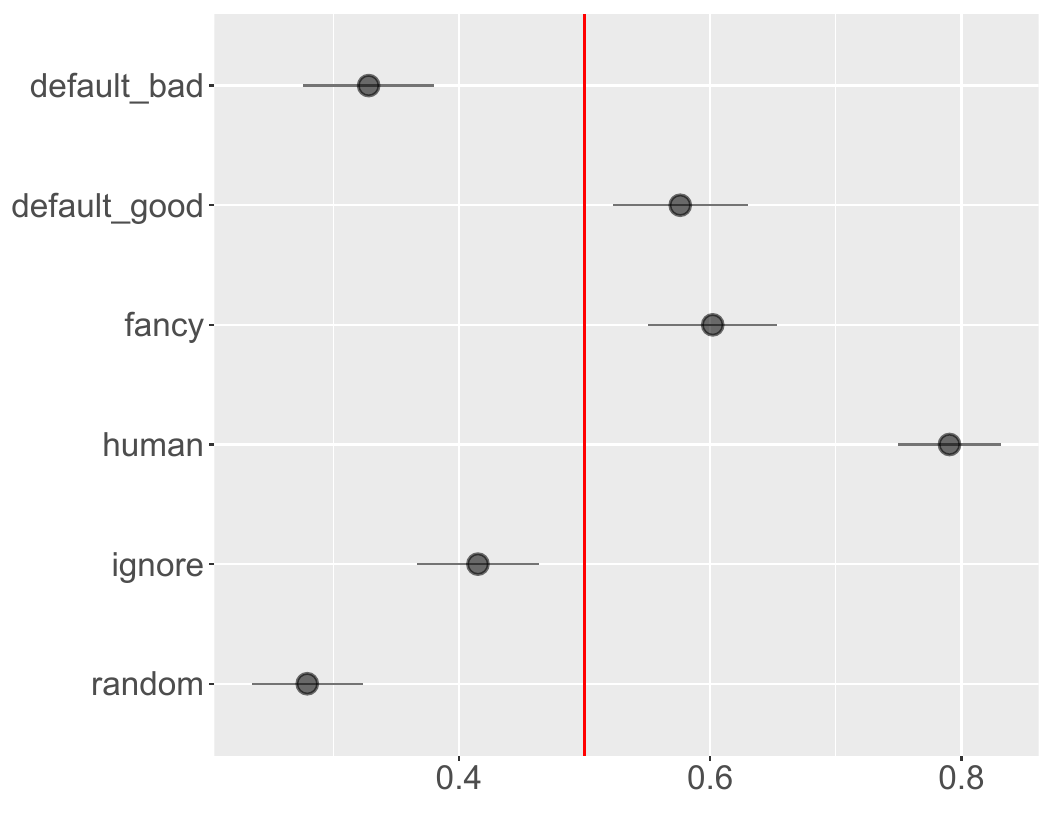}
    \end{center}
    \caption{ Marginal means for each multiplicity resolution technique. A score of 0.5 indicates the frequency the option would have been chosen if the selection is purely random, while a score greater (less) than 0.5 indicates an option is chosen more (less) frequently than would be expected by random chance.}
    \Description{The plot shows the marginal mean estimates for each multiplicity-resolution technique. We will report the lower bound, mean estimate, and upper bound for each. Default bad: 0.276, 0.328, 0.380. Default good: 0.522, 0.576, 0.630. Fancy: 0.551, 0.602, 0.653. Ignore: 0.367, 0.41, 0.464. Human: 0.749, 0.790, 0.832. Random: 0.235, 0.279, 0.323. }
    \label{fig:part2}
\end{wrapfigure}

\hTwoMain is significant since different multiplicity resolution techniques are selected at statistically significantly different rates (see \cref{fig:part2}). For instance, \texttt{human} is selected in a pairwise comparison at a rate of 0.795 (SE = 0.015, p < 0.0001) and \texttt{random} is selected at a rate of 0.279 (SE = 0.017, p < 0.0001). A post-hoc test (detailed results in appendix) reveals that the preference for \texttt{human}  is statistically stronger than all other options, while \texttt{random}  is significantly less preferred than all options other than \texttt{default-bad}.

\hTwoStakes is significant as well, and one of our four specific predictions holds: \texttt{human} is more strongly preferred when a task is high stakes. 
\hTwoFraming is significant, as is one of our four predictions: when the task is reward-based, \texttt{default-good} is more strongly preferred than when the task is punishment-based. 
\hTwoUnc is not significant.
Marginal mean estimates and details on t-test results are in the appendix.

\subsection{Exploratory analyses}\label{sec:res_explore}
We describe five exploratory results (labeled \explLikert-\explQual). 

\paragraph{\explLikert (Differences between different dimensions of fairness)}
While the set of 4 Likert items has high internal agreement for both the ``pre'' and ``post'' timestamps, the changes that occur with time are not distributed randomly across the four items.
Overall, there is not a significant difference between ``pre'' and ``post'' fairness scores (\hOneMain). 
But the first two Likert items, which ask about model outcomes (whether a system will be able to make good/fair decisions), see the scores decrease by an average of 0.122 points per item, while
the second two Likert items, which ask about the appropriateness of using the AI system, increase with time by an average of 0.131 points per item.
This difference is statistically significant according to the Mann-Whitney U-test (W=74118, p=0.0001). 

\paragraph{\explDuration (Survey duration's impact on change in fairness scores)}
We compute Spearman's rank correlation between survey duration and average change in fairness score. There a very weak positive correlation ($\rho = 0.137$, p=0.0002).

\paragraph{\explTrends (Trends in multiplicity solution preferences)}
We observe several trends that our hypotheses did not predict. 
(All results in the following paragraph are for t-tests based on marginal means estimates.) 
When stratifying tasks by stakes, \texttt{default-bad} is more preferred in low-stakes settings than in high-stakes settings (difference in marginal means -0.166, SE=0.0386, t=-4.307, p<$10^{-4}$). 
We hypothesized that \texttt{random} would be more preferred in low-stakes solutions (\hTwoStakes); this was not significant with the multiple hypothesis correction but shows a trend in the right direction (difference -0.079, SE=0.033, t=-2.411, p=0.016). 
The \texttt{default-bad} option is also more preferred in punishment settings than in reward settings (difference -0.183, SE=0.039, t=-4.692, p<$10^{-5}$), while \texttt{random} is more preferred in reward settings (difference 0.108, SE=0.033, t=3.31, p=0.0009). 
In high-uncertainty settings, \texttt{default-good} and \texttt{default-bad} are both more preferred than in low-uncertainty settings (difference 0.108, SE=0.044, t=2.441, p=0.015 and difference 0.089, SE=0.0387, t=2.310, p=0.021, respectively). 

\paragraph{\explCost (Role of cost)} A marginal-means analysis shows that \nocost is usually significantly preferred over \extracost (t-test result: difference   0.225, SE=0.016, t=13.91, p$<10^{-5}$). However, in three settings (tumor detection, care plan access, and job probation), the marginal means for \nocost and \extracost are not statistically different (see the appendix for complete data).

\paragraph{\explQual (Qualitative analysis of open-ended question)}  
We analyze the responses to the open-ended question as follows:
First, we exclude responses that are unclear or do not suggest a specific decision-making strategy.
Then, we categorize responses according to the resolution techniques from \cref{tab:part2_options} and an ``other'' category. 
We then review the responses in the ``other'' category and group similar responses into new categories.
If a response mentions multiple approaches, we count it in each category. 
Note that at this stage of the survey, participants had not seen the options in \cref{tab:part2_options}. We report general trends here; full statistics are available in the appendix.

The majority  of participants across all tasks mention \texttt{human}, including over 80\% of participants for job probation, care plan access, and tumor detection.
\texttt{default-good} is mentioned most frequently for tumor detection, tax audit, and beta app access, while \texttt{default-bad} is suggested for beta app access and flight upgrade.
\texttt{random} is suggested frequently for flight upgrade and occasionally for tax audit and flight rebooking. 
Participants also frequently suggested specific rules they would implement or indicated that a more complex model could be built to resolve disagreements. 
This type of answer was especially common for flight upgrades and rebooking, review moderation, and beta app access. 
Finally, a few participants suggested involving the decision subject directly, e.g., asking a flight passenger how they feel about being rebooked or upgraded. This option was mentioned by at least one respondent for flight rebooking, flight upgrade, and care plan access.

\section{Discussion}

\subsection{The standard ML approach is unfair (\texttt{ignore} is disliked)}
Our most impactful finding is that \texttt{ignore} is generally not preferred. 

\implication{Standard ML workflows need to adapt to consider multiplicity, such as by increasing transparency when multiplicity impacts a decision.}

Depending on the target audience, interventions to increase transparency could take different forms.
For instance, common machine-learning packages could alert  developers when multiplicity exists (e.g., by automatically performing a bootstrapping analysis in the background). 
Model documentation should include an analysis of what proportion of test samples are subject to receive a conflicting prediction under multiplicity, so that anyone who decides to use or adapt the model in the future is aware.
In public communications about a model, the fraction of inputs (based on validation data) that are subject to receive a different prediction under multiplicity could be reported.

\implication{Model developers should be able to justify how multiplicity is handled, whether this is by deferring to a human decision-maker, employing randomization, using ensemble methods, or something else.}

It is possible that decision subjects would be okay with using a single model, and knowingly ignoring multiplicity, if this decision is justified. But, just as model developers need to justify decisions such as feature selection and statistical fairness trade-offs, they should justify why the existence of multiplicity does not pose an existential threat to their project. In particular, they should consider potential harms to individuals, e.g., whether the model's use will be widespread and lead to systemic exclusion of some people.

\subsection{Is multiplicity an inherent fairness risk?}
Our results show that participants' fairness perceptions did not decrease when multiplicity was present (i.e., \hOneMain was not supported). This contradicts work in ML and philosophy that assumes and argues that multiplicity poses a fairness risk (see \cref{sec:mult}). Since \hOneMain was not supported by our data, either (1) the experts are incorrect about the fairness risks of multiplicity, (2) lay stakeholders do not see fairness risks of multiplicity due to low technical literacy, or (3) our study design was insufficient to capture stakeholders' true views on multiplicity. We discuss (3) when describing our work's limitations (\cref{sec:disc_limitations}), so we will overview (1) and (2) here.

If (1), the fairness risks of multiplicity summarized in \cref{sec:mult} must paint an incomplete picture.
Perhaps the philosophical view that only systematic arbitrariness is a fairness risk~\cite{creel2022leviathan} is correct.
Alternatively, maybe multiplicity poses a \emph{theoretical} fairness risk, but one that is is either too small to matter in practice, or that is overpowered by the benefits of using ML. 
Participants may feel as if the ML system is a closed box, and thus that its decisions are impenetrable. This perception would, understandably, lead participants to feel disenfranchised in their interactions with the ML model. From this perspective, multiplicity is not the primary reason why a model's usage could be unfair.

\implication{Theoretical researchers should articulate how the fairness risks of multiplicity relate to other fairness issues in ML -- such as accountability and recourse -- to help clarify when, exactly, multiplicity poses a fairness risk. Future research about the perceived fairness of multiplicity should do more to access the ``why:'' what are the driving forces in fairness perceptions?}

If (2), then \hOneMain was not supported because the stakeholders' technical literacy was not high enough to realize that multiplicity poses the procedural fairness risks suggested by computer scientists and philosophers. As ML technology powers more and more decisions that used to be made by humans, it is important that stakeholders are able to recognize its limitations to avoid 
inappropriate reliance on and trust in ML systems~\cite{long2020what}.
One way that reliance and trust can be calibrated is through improving technical/AI literacy~\cite{chiang2022exploring}. 
We discuss how our survey design could be modified to promote better understanding in \cref{sec:disc_limitations}.
But also, technical literacy can be built gradually over time by exposure to news stories and public discourse~\cite{nguyen2024news}. 

\implication{We encourage people working at the intersection of technology and communication to discuss multiplicity and to take a critical lens, more broadly, when describing new ML systems in socially-impactful domains.}

\subsection{Randomization is disliked}
Scholars in ML and philosophy typically suggest using a more complex model (e.g., via ensembles)~\cite{cooperArbitrariness,roth2022reconciling} or randomization~\cite{creel2022leviathan,jain2024position}. 
Participants in the survey expressed comfort with \texttt{fancy} -- choosing this option more often than would be expected by random chance -- but strongly disliked \texttt{random} (\hTwoMain). 
This contradicts the established philosophical view that lotteries are fair in situations where a scarce resource must be allocated~\cite{broome1990fairness,sher1980makes}. 
A possible discrepancy between our setup and a run-of-the-mill lottery is the inclusion of machine learning in the decision-making process. Perhaps randomization via a lottery is familiar, but the addition of machine learning suggests that there must be some way to ``learn'' the best allocation, thereby making randomization too crude of a tool. 

\implication{More research should be conducted into why \texttt{random} is disliked, even though lotteries are generally viewed as fair. Researchers can investigate whether providing additional education about machine learning (e.g., how models can make mistakes and that there is often not a single, optimal solution) improves comfort with \texttt{random}.}

\subsection{Participants prefer human intervention, but is that scalable?}
Participants generally prefer \texttt{human} over other multiplicity-resolution techniques (\hTwoMain). However, when large proportions of predictions can vary under multiplicity, deferring to a human decision-maker is difficult to operationalize and negates cost and time savings associated with using ML.
In these cases, one option is to simply forgo using ML at all to avoid the appearance of automation and the expense of developing a system that only handles a small number of cases. (Indeed, it may not be scientifically valid to use machine learning when many predictions vary under multiplicity, as has been argued for ``predictive optimization'' tasks~\cite{wang2024against}.) 
Instead, decision-makers could rely on an interpretable, publicly-available set of guidelines. Alternatively, model developers could use a strategy that combines \texttt{fancy} and \texttt{human}. Recall that \texttt{fancy} (e.g., using ensemble methods) is favored by ML researchers as a technique to reduce multiplicity. Model developers could use ensembles or more complex model architectures to reduce multiplicity, and then defer to a human decision maker only for the smaller number of cases whose predictions remain ambiguous. In our experiments, participants were generally comfortable with \texttt{fancy}, so while further research is needed to confirm that a combination \texttt{fancy} and \texttt{human} approach would be viewed as fair, we are optimistic about that possibility.

\implication{Model developers may not be able to align with stakeholders' preferences for \texttt{human} in all cases, but should consider whether other, more interpretable decision-making rules are more appropriate when multiplicity is present. Alternatively, using ensemble methods to reduce multiplicity, and then deferring to a human on the smaller number of ambiguous cases, could strike a good balance.}

\subsection{Impact of task choice}
Preferences for multiplicity resolution techniques varied across tasks. 
\explTrends shows that \texttt{random} is more preferred in general for low-stakes scenarios, but also for one high-stakes scenario (tax audit). 
We hypothesize that this is because decisions about whom to audit are seen as more impersonal than the other high-stakes scenarios. 
In medical and employment domains, decision subjects typically have a personal relationship with their doctor or manager, and thus trust that they will have the ability to appeal to human emotion and get an explanation of the decision. 
By contrast, decisions by the IRS (the US tax agency) about whom to audit likely already feel random to many individuals. 
Conversely, \texttt{human} is more strongly preferred in high-stakes settings where a personal connection is the norm (\hTwoStakes \& \explTrends). 
We also want to call attention to how 
\texttt{ignore} is usually preferred to \texttt{random} (\hTwoMain). 
From the perspective of a decision subject who interacts with a model a single time,  there is not a mathematical difference between \texttt{ignore} and \texttt{random} because the personal outcome is unknown and (relatively) unintelligent in each. 
(By contrast, \texttt{default-good} and \texttt{default-bad} give a known outcome, and people may assume that they can appeal to a human decision maker to get the correct/favorable outcome with \texttt{human} and that  \texttt{fancy} will be smart enough to give the correct decision.)
So, individuals must either be imagining repeated encounters with a system or just view having a (presumably) deterministic model as providing benefits such as accountability and predictability. 

\implication{Model developers should involve participants in the design process to figure out how they prefer that multiplicity be handled for the specific task.}

\subsection{Limitations and opportunities for future work}\label{sec:disc_limitations}
We conclude our discussion with an overview of limitations and how they could be addressed in future work.

\paragraph{Cognitive load for participants}
A limitation of this work is that multiplicity is not an easy concept to grasp, especially for people without machine learning experience. We attempted to explain it clearly and in simple terms and iterated on the survey text through multiple pilot iterations. 
Our results do not show significant differences in the responses for participants with or without ML experience.
We attempted to check that participants understood multiplicity through an initial comprehension check, but some participants may have gotten lucky and passed this question by chance, only looked for the correct answer rather than engaging critically with the reading, or may have forgotten details when answering the remaining survey questions. 

\paragraph{Unknown role of other task-related factors}
To counteract the cognitive load required to reason about multiplicity, we tried to make the other details about the tasks and models as simple as possible, including by providing only high-level details of the AI system's functionality and context. Although we tried to select tasks that are broadly relevant, some participants may not have seen the relevance to their own lives or imagined themselves as the decision subjects. Some participants may have felt like they did not have enough information to evaluate the system's fairness, and the presence of multiplicity may have seemed insignificant relative to other unknown factors. 
In particular, we did not include any explanations of how the models made their decisions to keep the setup simpler. Informational transparency is a component of overall fairness~\cite{colquitt2001dimensionality}, though, so future work could explore how different explanations of both model behavior and of multiplicity itself impact fairness perceptions. 
Likewise, we do not include personal impact -- i.e., whether the prediction a model makes on a specific decision subject varies under multiplicity -- as a factor due to the experiment's complexity. We hypothesize -- and future work should check -- that if participants are told that the decision the model makes for them varies under multiplicity, fairness perceptions would decrease, mirroring work on outcome favorability~\cite{wang2020factors}. 
To combat issues with the large number of unknowns inherent to hypothetical situations, future research could be situated in a real-world task, where explaining more details about the model would not necessitate a large cognitive load due to baseline familiarity with an existing algorithm. Existing research also shows that people behave differently in hypothetical settings~\cite{feldmanhall2012we}, again highlighting the need for additional research situated in a real use case.

\paragraph{Unknown role of other personal factors}
We did not find significant trends between demographic groups for any of the factors we recorded. However, our analysis was not powerful enough to detect how any intersectional identities (e.g., gender and education level) may impact perceptions of multiplicity. Furthermore, our collection of some demographic attributes, especially experience with AI, may have been too coarse to see meaningful trends. Future work could consider contrasting specific participant groups (e.g., ML engineers versus adults with no ML experience) to gain a better sense of how fairness perceptions vary with ML experience.  

\paragraph{Addressing the ``why'' behind multiplicity preferences}
Our results indicate that stakeholders have strong preferences for how multiplicity is handled; however, our experiment does not determine why these preferences exist.
In particular, the discrepancy between philosophical arguments for the fairness of \texttt{random} and the strong dislike that survey participants expressed towards \texttt{random} merits more attention. 
Is there truly a mismatch between what is theoretically fair based on philosophical principles and what people perceive as fair? 
Or is there a breakdown in the communication, e.g., would participants actually be comfortable with randomization if more details about the model and its operationalization were provided? 
Future studies could vary the amount of detail provided to see whether attitudes towards randomization change.

\paragraph{Task choice and characterization} We consider tasks that differ along three axes, but there may be alternate ways to characterize tasks. It is also not necessarily straightforward to introduce new tasks into our classification system, particularly for the continuous scale of task stakes and task uncertainty. Our analysis also lost detail about the task characteristics in the conjoint analysis, as this framework does not support continuous factors. 
Future work could continue to study how to systematically characterize tasks to promote reproducible and generalizable research.

 \paragraph{US-centric perspective} We conducted our survey with US-based crowdworkers. It is known that fairness perceptions vary cross-culturally~\cite{berman1985cross,kim2007forming}, as do attitudes towards uncertainty~\cite{hofstede1984culture}.
It therefore seems likely that attitudes towards the fairness of multiplicity in ML would vary across cultures, too, which future work could explore.

\section{Conclusions}

We conducted a study (n=357) of crowdworkers to determine how lay stakeholders (1) perceive the fairness of ML models when multiplicity impacts decisions, and (2) prefer that this multiplicity be resolved.
Our results indicate that stakeholders have strong preferences about how multiplicity should be resolved. 
In particular, they express a preference for human decision makers, and a dispreference for both ignoring multiplicity (arbitrarily using a single model) and randomizing between different models' outputs. 
These findings counter not only the established ML practice of looking for a single well-performing model, but also philosophical arguments that randomization is the fairest way to deal with multiplicity.
Despite these strong opinions about resolving multiplicity, the presence of multiplicity did not impact how fair the participants thought the models were, which is a discrepancy that future work will need to address.
One promising avenue is to consider how perceptions of informational fairness relating to multiplicity, e.g., the use of explanations, intersects with the procedural components of fairness that we investigated.
Ultimately, a greater understanding of when ML design choices -- including how multiplicity is resolved -- are perceived as fair can help guide model developers to create technology that aligns with stakeholders' values.

\begin{acks}
The authors thank members of the MadPL group for serving as pilot testers for initial survey iterations, and Michael Coblenz for his comments on an earlier draft. 
This work was conducted with the support of NSF grants 1918211 and 2446711.
\end{acks}

\newpage

\bibliographystyle{ACM-Reference-Format}
\bibliography{sources}

\newpage
\appendix
\section{Preliminary study}

\subsection{Survey materials}
We ask the following questions (on 5-point Likert scales from strongly disagree to strongly agree) to gauge task stakes (S), task uncertainty (C), and task clarity and realistic-ness (R). Questions are flagged with (*) to indicate that the phrasing changes with the scenario to provide more detail (e.g., for the task of loan approval ``in the positive direction'' would become ``approves a loan for someone who will not be able to repay it'', since getting a loan is the positive outcome from the perspective of the decision subject). 

\begin{itemize}
    \item (S) This scenario describes a high-stakes situation.
    \item (S) It is a big problem if the system makes a mistake in the positive direction.*
    \item (S) It is a big problem if the system makes a mistake in the negative direction.*
    \item (C) For each decision the system makes, it is objectively right or wrong (i.e., there is not much ambiguity in decisions).
    \item (C) If the AI system is designed well, I think it would be able to make better decisions than a human expert. 
    \item (R) Is this scenario described clearly?
    \item (R) Is this scenario realistic?
\end{itemize}

\subsection{Tasks and results}
\Cref{tab:all_tasks} shows all tasks we considered in the preliminary study, along with their final stakes and uncertainty scores. 

\begin{table*}[t]
\caption{Summary of tasks and their stakes, uncertainty, and framing as determined by the preliminary study. Tasks marked with (*) were excluded from the main study because participants indicated they were unclear and/or unrealistic. The final column ``Use'' indicates whether we use the task in the main study.} 
\label{tab:all_tasks}
\begin{tabular}{ll|rrcc}\toprule
Domain & Task & Stakes & Uncertainty & Framing & Use    \\\midrule
Healthcare & Access to a preventative health care plan & 0.42 & 0.91 & Reward  & Yes   \\
Business & Job probation based on employee monitoring & 0.72 & 1.12 & Punishment & Yes \\ 
Healthcare & Tumor detection from medical imaging & 1.61 & -1.37  & Reward  & Yes   \\
Government & Tax audit & 0.44 & -1.34  & Punishment & Yes \\
Travel & Airline seat upgrade & -1.94  & 1.18 & Reward  & Yes   \\
Travel & Rebooking passengers on an over-full flight & -0.27  & 0.50 & Punishment & Yes \\
Business & Customers get beta access to an app with discounts & -1.58  & -0.21  & Reward   & Yes  \\
Business & Removing online reviews that are spam or fake & -1.26  & -0.16  & Punishment & Yes \\
Education & College admissions decisions* & 0.40 & 1.13 & Reward & No \\
Education & Automated plagiarism detection & -0.16 & -1.47 & Punishment & No \\
Healthcare & Discharge from hospital* & 1.41 & 0.50 & Punishment & No \\
Business & Hiring based on resume screening & -0.18 & 1.04& Reward & No \\
Business & Approving online reviews for publication & -1.15 & -0.06 & Reward & No \\
Banking & Detecting fraudulent credit card transactions & 0.44 & -1.19 & Reward & No \\
Banking & Approving loan applications & 0.49 & -0.01 & Reward & No\\
Government & Determining benefits eligibility* & 0.92 & -0.31 & Reward & No\\
Insurance & Adding an insurance surcharge if high-risk & 0.11 & -0.12 & Punishment & No \\
Insurance & Given an insurance discount if low-risk & -0.40 & -0.15 & Reward & No \\
\bottomrule     
\end{tabular}
\end{table*}

\section{Complete survey materials}
\subsection{Tasks} 
This section contains the full text used to describe each task at the beginning of Part 1.
\paragraph{Care plan} A hospital plans to use an AI system to identify patients who qualify for a preventative care plan, which includes access to educational materials and more frequent doctor check-ins. The AI will base its decisions on the patient's health and past disease progression patterns. Doctors can override the AI's decisions if needed.
The hospital checks that the AI works well by seeing how often it agrees with doctors' choices. The AI and the doctors agree most of the time.

\paragraph{Job probation}
A company is planning to use an AI system to identify underperforming employees who need job probation. The AI will make its decisions based on performance metrics and work history. All flagged employees will be placed on probation, but they will not be demoted or fired unless a human review of their probationary period is negative.
The company checks that the AI works well by seeing how often it agrees with human managers' choices. The AI and the managers agree most of the time.

\paragraph{Tumor detection}
A hospital plans to use AI system to look at CT scans to find potential cancerous tumors. If the AI detects signs of a tumor, the patient will receive additional tests. However, if no tumor is detected, additional tests will not be conducted. Doctors can override the AI's decision if needed.
The hospital checks that the AI works well by seeing how often it agrees with doctors' choices. The AI and the doctors agree most of the time.

\paragraph{Tax audit}
A government tax agency is planning to use an AI system to detect potential tax fraud. Everyone flagged by the AI will be audited, and audits will occur solely based on AI flags. However, taxpayers will only face penalties if a human investigation confirms the fraud.
The tax agency checks that the AI works well by seeing how often it agrees with human experts' choices of who to audit. The AI and the human experts agree most of the time.

\paragraph{Upgrade}
An airline is planning to use an AI system to decide which passengers get upgrades to empty first-class seats. The AI will make its decisions based on factors like passenger loyalty status, ticket price, and check-in time. The AI's decisions are final, but passengers can decline the upgrade and keep their original seat if they prefer.
The airline checks that the AI works well by seeing how often it agrees with trained airline employees' choices. The AI and the airline employees agree most of the time.

\paragraph{Rebook}
An airline wants to use an AI system to manage overbooked flights. The AI will re-route passengers based on their destination, personal details (like age and group size), and booking details. While the AI's decisions are final, passengers can opt for a refund or select an alternative re-routing plan if they are unhappy with the AI's choice.
The airline checks that the AI works well by seeing how often it agrees with trained airline employees' choices. The AI and the employees agree most of the time.

\paragraph{App}
A company is planning use an AI system to decide which customers get beta access to their new app, which offers exclusive deals and coupons. The AI will make its decisions based on expected financial gain for the company based on the customers’ past spending habits. The decisions about which customers get early access are final until the app is available to the public.
The company checks that the AI works well by seeing how often it agrees with human marketing experts' choices. The AI and the marketing experts agree most of the time.

\paragraph{Reviews}
A website where people post reviews of businesses is planning to use an AI system to detect and remove fake reviews, e.g., posts by bots or by people who have not actually visited a business. Once a review is deleted, it cannot be recovered, but the person who wrote it can re-post a similar review if they want.
The review website checks that the AI works well by seeing how often it agrees with trained human moderators' choices. The AI and the moderators agree most of the time.

\subsection{Multiplicity education}
We provide the following text to participants at the beginning of the survey:

\begin{figure}
    \centering
    \includegraphics[width=0.75\linewidth, trim=0cm 6cm 0cm 6cm]{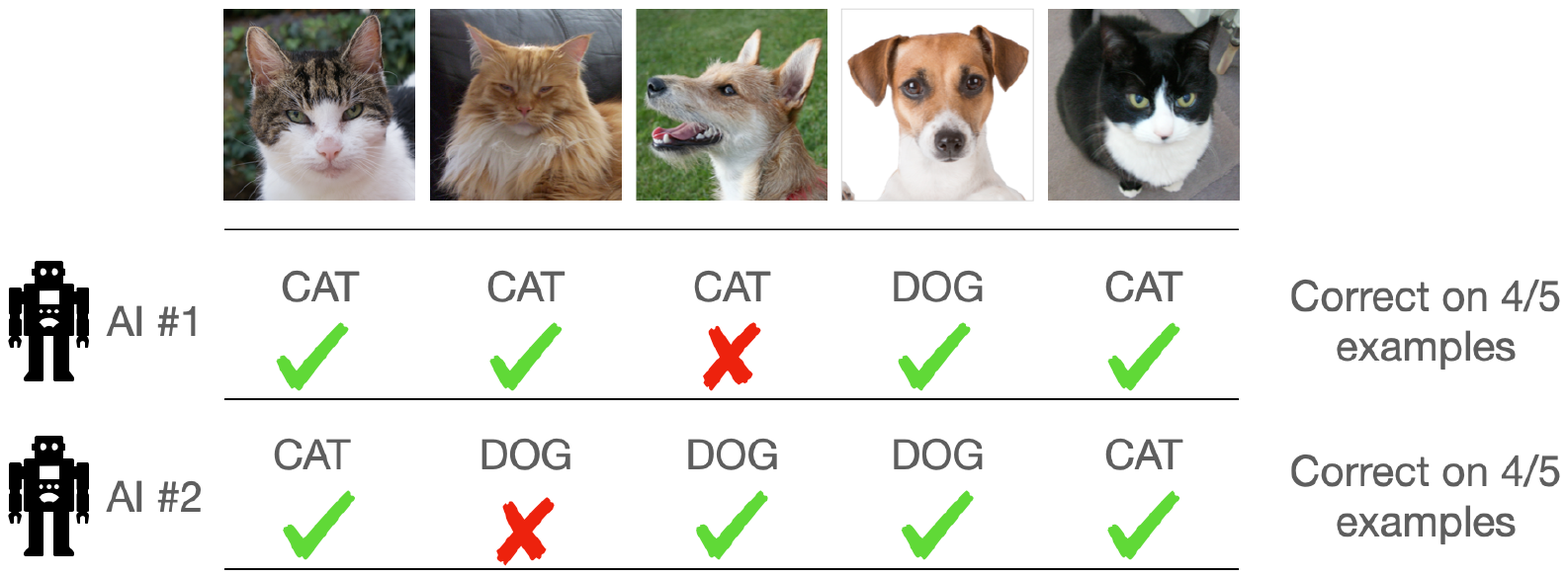}
    \caption{Image that we include with the multiplicity education.}
     \Description{The image shows a table with 5 columns and two rows. The column headers are pictures, each of a cat or a dog. The two rows are labels AI\#1 and AI\#2. The table cells indicate what the AI models predict as the label for each image. Each cell includes the prediction (i.e., the word ``cat'' or the word ``dog'') as well as a green checkmark, in the case of a correct prediction, or a red X in the case of an incorrect prediction. AI\#1 is correct on 4/5 examples (it identifies one dog (column 3) as a cat). AI \#2 is also correct on 4/5 examples, but identifies one cat (column 2) as a dog. Text to the right of each row points out that both AIs are correct on 4/5 examples.}
    \label{fig:educ_fig}
\end{figure}

\begin{quote}
   When building Artificial Intelligence (AI), choosing the best model is crucial. AI developers consider a range of AI models and choose the one with the best performance. For example, if the goal is to distinguish between cats and dogs, developers will test different AIs using a set of images and choose the AI that correctly identifies the most pictures.
\end{quote}

\begin{quote}
   In the example below, AI \#1 correctly identifies 4 out of 5 images. Suppose this is the highest-performing AI model available.
\end{quote}

\begin{quote}\textit{We display the top half of \cref{fig:educ_fig} (i.e., images and AI \#1 only).}\end{quote}

\begin{quote}
    Sometimes, however, multiple AI models perform equally well but differ in their specific errors. This is called model multiplicity. For example, AI \#2 also correctly identifies 4 out of 5 images but makes a different mistake than AI \#1.
\end{quote}

\begin{quote}\textit{We display \cref{fig:educ_fig}.}
\end{quote}

\begin{quote}
    AI \#1 and AI \#2 are both well-suited for classifying cat and dog images. If no other AI performs better, developers must choose between AI \#1 and AI \#2.
\end{quote}

\begin{quote}
    Model multiplicity is common in AI: multiple models can perform equally well overall but make different mistakes. In some settings, like predicting cats and dogs, model multiplicity is inevitable but choosing between models is low-stakes. In other settings, choosing between models is more challenging, such as when the model predictions directly affect people.
\end{quote}

Then, we ask the comprehension question ``What does model multiplicity refer to?'' with these choices: (a) It refers to a model making multiple correct predictions, (b) It means there are several models that perform equally well but makes different mistakes, (c) It describes an AI model that performs better than all other models, and (d) It refers to an AI model that can perform multiple task simultaneously.

\subsection{Part 1: Fairness questions}\label{sec:app_fairness_qs}
We use two questions relating to distributive fairness, and two relating to procedural fairness. Each question is on a 7-point Likert scale (``Strongly disagree'' to ``Strongly agree''). The exact phrasing of each question is tailed to each scenario to be more descriptive (e.g., replacing the word ``entity'' with the context-appropriate word).
\begin{itemize}
    \item (Distributive) I think the AI will make good decisions.
    \item (Distributive) I think the AI will make fair decisions.
    \item (Procedural) I think it is appropriate for the entity to use AI in this context.
    \item (Procedural) I would feel comfortable if the entity used AI to make a decision about me.
\end{itemize}

\subsection{Part 1: Critical-thinking questions}
We ask two critical-thinking questions midway through Part 1 of the survey to encourage participants to slow down and better absorb the material. The first question is multiple-choice, asking which of four statements is true: (a) The original AI model is the only good way to do [task], (b) The new AI and human experts always choose the same people, (c) Multiple models are equally good at [task], or (d) The original AI and new AI never agree on [task]. The correct answer is (c) and this is visible to participants as soon as they make any selection. The open-ended question asks ``If the original AI and the new AI disagree about whether or not to [predict + or -], how should the [decision-making entity] decide what to do?'' Respondents much enter at least 8 characters in their answer, but apart from that requirement there is no in-survey validation of the answer.

\subsection{Part 2: Costs}
\Cref{tab:costs} shows the (non-zero) cost that we use for each task.

\begin{table*}[t]\caption{The non-zero cost for each task.}\label{tab:costs}
\begin{tabular}{l|l}\toprule
Task & Cost    \\\midrule
Care plan &  15 minute longer wait time at appointments to see the doctor \\
Job probation & Employees will spend 15 extra minutes per week filling out paperwork about their work \\ 
Tumor detection & 15 minute longer wait time at appointments to see the doctor \\
Tax audit & There will be an additional \$5 fee to file taxes \\
Upgrade & Airline tickets will cost \$5 more \\
Rebook & Airline tickets will cost \$5 more \\
App & The shipping time on all products will be 2 days longer \\
Reviews & The website will have to host more ads, including pop-up ads \\
\bottomrule     
\end{tabular}
\end{table*}

\section{Alternate statistical analyses}
We describe two alternate statistical analyses for the Part 1 data (change in fairness perceptions after learning about multiplicity). As stated in \cref{sec:method_stats}, we preregistered our study with a plan to use linear mixed models. Those results are in the main text. We include two alternate analyses here for two reasons: first, there is disagreement in the statistical community about what type of analysis is appropriate for Likert-scale data~\cite{blanca2017non,carifio2008resolving,jamieson2004likert,kuzon1996seven,norman2010likert}. Including multiple options serves to ``future-proof'' our analysis in the case of future statistical consensus. Second, while preregistration is one tool to guard against ``researcher degrees of freedom''~\cite{simmons2011false}, another option is a multiverse analysis~\cite{steegen2016increasing}. In a true multiverse analysis, researchers present the outcomes of all feasible statistical analyses on their data. Then, if the same conclusions hold for all or most of the analyses, they can be considered more robust. We present a partial multiverse analysis. 

\subsection{ART-ANOVA}
Aligned-Rank-Transform (ART) ANOVA~\cite{wobbrock2011aligned} is a procedure that transforms ordinal (e.g., Likert-scale) data to fit the ANOVA assumptions.
We fit our ART-ANOVA model and perform post-hoc contrast tests~\cite{elkin2021aligned} using the ARTool package in R~\cite{artool}. Our formula includes the factors time, task stakes (binary), task uncertainty (binary), and task framing and all possible interactions between factors. We also include the participant ID to account for within-subjects factor.
We test \hOneMain using a post-hoc contrast test with ``time'' as the factor of interest. We test \hOneStakes, \hOneFraming, and \hOneUnc with a post-hoc contrast test of the interaction term between ``time'' and the relevant task characteristic (with the ``Interaction'' flag set to true). 

\paragraph{Results}
Our ART-ANOVA model indicates that the following factors have significance: task stakes (F=35.2, Df=1, p$<10^{-8}$), task framing (F=5.0, Df=1, p=0.026), and the interaction term between task stakes and task uncertainty (F=7.9, Df=1, p=0.005). As was the case when using LMMs in the analysis, none of the hypotheses \hOneMain-\hOneUnc was significant. 

\subsection{Non-parametric tests}
The Mann-Whitney U-test~\cite{mann1947test} is another way we can evaluate \hOneMain-\hOneFraming. This test is non-parametric, so does not require any assumptions about the data's normality or distribution. 
We perform the tests using the \texttt{wilcox.test} function in R. For \hOneMain, we test whether the average fairness perception is different as stratified by ``time''. For \hOneStakes-\hOneUnc, we compute the change in score over time for each sample, and then conduct a Mann-Whitney U test on the data stratified by the relevant task characteristic.

\paragraph{Results} As is the case with using LMMs or ART-ANOVA, none of the hypotheses \hOneMain-\hOneUnc is significant. 

\section{Additional results (Fairness perceptions)}

\subsection{Task choice and baseline fairness perceptions}

\Cref{tab:baseline_fairness} shows the fairness perceptions for each task at the ``pre'' timestamp. These values should be interpreted as a baseline for how fair participants think it is to ML in each scenario. 
We observe that high-stakes tasks have lower fairness scores, as do punishment-framed tasks. 

\subsection{Demographic factors}
We consider binary gender, integer age, binary experience with AI (defined as whether the participant currently uses AI in their work or school), and binary education (whether the participant has a bachelor's degree or higher). 
There is no significant difference in initial fairness perceptions based on gender (t-test p=0.822), age (linear regression p=0.184), experience with AI (t-test p=0.088), or education (t-test p=0.328). 
There is no significant different in change in fairness scores after learning about multiplicity for gender (t-test p=0.921), age (linear regression p=0.944), AI experience (t-test p=0.460), or education (t-test p=0.413).

\begin{table}[th] \caption{ The average fairness score (on the questions listed in \cref{sec:app_fairness_qs}) at the ``pre'' timestamp for each task. Scores are from 7-point Likert scales; re-coded as -3 (Strongly disagree that the model is fair) to 3 (Strongly agree that the model is fair). We report the mean with the interquartile range in parentheses. We also indicates stakes (binary low/high), uncertainty (binary low/high), and framing for reference.}\label{tab:baseline_fairness}
\begin{tabular}{l|rlll}\toprule
Task & Score & Stakes & Unc. & Framing    \\\midrule
Care plan &  0.26 (-0.75, 1.13)  & High & High & Rew. \\
Job probation & -0.38 (-1.50, 0.88) & High & High & Pun. \\ 
Tumor detection & 0.12 (-0.75, 1.00) & High & Low & Rew.\\
Tax audit & 0.19 (-0.25, 1.00) & High & Low & Pun. \\
Upgrade & 1.01 ( 0.75, 1.75) & Low & High & Rew.\\
Rebook & 0.83 ( 0.50, 1.75) & Low & High & Pun. \\
App &  0.78 ( 0.00, 1.50) & Low & Low & Rew. \\
Reviews & 0.55 (-0.25, 1.50) & Low & Low & Pun. \\
\bottomrule     
\end{tabular}
\end{table}
For high-stakes tasks, the mean fairness score is 0.05 on a 7-pt scale (IQR -0.81 to 1.0) and for low-stakes tasks it is 0.79 (IQR 0.25 to 1.5). 
This difference is statistically significant according to a Mann-Whitney U-test (W = 43476, p $<10^{-13}$). 
Punishment-framed tasks have a median fairness score of 0.30 (IQR -0.5 to 1.25), while for reward-framed tasks it is 0.55 (IQR 0.0 to 1.5). This difference is also statistically significant according to a Mann-Whitney U-test (W = 71016, p = 0.012). 
Task uncertainty does not have an impact on fairness score according to a Mann-Whitney U-test (W = 64358, p = 0.918).

\section{Additional results (Multiplicity resolution)}

\subsection{Details on testing \hTwoMain-\hTwoUnc}
\Cref{tab:part2_posthoc} shows the effect sizes and significant for the post-hoc analyses conducted for \hTwoMain-\hTwoUnc (\cref{sec:res_hypo}). We see that all effects except two are in the expected direction, and, using the Holm-Bonferroni correction, 10 of the 18 tests are significant.

\subsection{Qualitative results}
\Cref{tab:qual} shows the results of the open-ended question, where participants were asked how to resolve multiplicity prior to seeing the techniques that we proposed. The table contains 10 response categories and tallies the number of participants whose answer referenced that category for each of the 8 tasks. We did our best to interpret the responses, however, in some cases, we could not determine what the participant meant so we marked it as ``unclear''. Most of these responses were on-topic and were a clear attempt to answer the question. 

\begin{table*}[t]\caption{Results of post-hoc t-tests for evaluating \hTwoMain-\hTwoUnc. The ``diff'' column is the difference in marginal means between the two listed factors (first 9 rows) or between the two test conditions for a single factor (remaining rows). (*) indicates that an effect is in the opposite direction as expected.}\label{tab:part2_posthoc}
\begin{tabular}{llrrrrl}\toprule
Hypothesis                & Factor          & Diff & SE & t-stat & p-val & Sig? \\\midrule
\hTwoMain & \texttt{human} v. \texttt{fancy} & 0.188 & 0.025 & 7.563 & <0.0001 & Yes  \\
\hTwoMain & \texttt{human} v. \texttt{default-good} & 0.214  & 0.026 & 8.382 &  <0.0001 & Yes    \\
\hTwoMain & \texttt{human} v. \texttt{default-bad}  & 0.462 & 0.025 & 18.340 & <0.0001 & Yes \\
\hTwoMain & \texttt{human} v. \texttt{ignore} & 0.375 & 0.024 & 15.594 & <0.0001 & Yes \\
\hTwoMain & \texttt{human} v.  \texttt{random} & 0.511 & 0.023 & 22.407 & <0.0001 & Yes \\
\hTwoMain &  \texttt{random} v. \texttt{fancy} & -0.323 & 0.026 & -12.592 & <0.0001 & Yes \\
\hTwoMain &  \texttt{random} v. \texttt{default-good} & -0.297 & 0.026 & -11.304 & <0.0001 & Yes \\
\hTwoMain &  \texttt{random} v. \texttt{default-bad} & -0.049 & 0.026 & -1.887 & 0.0593 & No \\
\hTwoMain &  \texttt{random} v. \texttt{ignore} & -0.136 & 0.025 & -5.477 & <0.0001 & Yes \\
\hTwoStakes & \texttt{human} & 0.097 & 0.030 & 3.211 & 0.0013 & Yes \\
\hTwoStakes & \texttt{fancy} & 0.043 & 0.038 & 1.128 & 0.2596 & No \\
\hTwoStakes & \texttt{ignore} & -0.089 & 0.036 & -2.468 & 0.0137 & No \\
\hTwoStakes & \texttt{random} & -0.079 & 0.033 & -2.411 & 0.0160 & No \\
\hTwoFraming & \texttt{human}* & -0.004 & 0.031 & -0.133 & 0.8943 & No \\
\hTwoFraming & \texttt{default-good} & 0.016 & 0.040 & 3.965 & < 0.0001 & Yes \\
\hTwoFraming & \texttt{fancy} & -0.059 & 0.039 & -1.510 & 0.1313 & No \\
\hTwoUnc & \texttt{random} & 0.010 & 0.033 & 0.300 & 0.7641 & No \\
\hTwoUnc & \texttt{fancy}* & 0.015 & 0.039 & 0.389 & 0.6970 & No \\
\bottomrule 
\end{tabular} 
\end{table*}

\newcommand*\rot{\rotatebox{45}}

\begin{table*}[t]
\caption{Results of our qualitative analysis. The number in the header indicates how many participants were assigned to that group. If a response suggests multiple techniques, we count it in all categories. (*) Unclear indicates we could not interpret the response, while Not sure means that the participant wrote that they did not know how to resolve the multiplicity.} 
\label{tab:qual}
\begin{tabular}{l|rrrrrrrrr}\toprule
& \multirow{2}{*}{Care plan} & \multicolumn{1}{c}{Job} & \multicolumn{1}{c}{Tumor} & \multirow{2}{*}{Tax audit} & \multirow{2}{*}{Upgrade}&\multirow{2}{*}{Rebook} & \multirow{2}{*}{App} & \multirow{2}{*}{Reviews} & \multirow{2}{*}{Total} \\
& & probation & detection & & & & & \\
& $n$=43 & $n$=43 & $n$=45 & $n$=45 & $n$=46 & $n$=45 & $n$=46 & $n$=45 & $n$=358 \\\midrule
\texttt{human}        & 35 & 35 & 38 & 23 & 22 & 28 & 28 & 35 & 244 \\
\texttt{default-good} &  1 &  1 &  7 &  7 &    &    &  3 &  2 &  21 \\
\texttt{ignore}       &  1 &    &    &  1 &  5 &  5 &  5 &  2 &  19 \\
\texttt{fancy}        &  1 &  2 &    &    &  5 &  1 &  3 &  2 &  14 \\
\texttt{random}       &    &    &    &  2 &  5 &  2 &  1 &    &  10 \\
\texttt{default-bad}  &  1 &  1 &    &  3 &  1 &    &  3 &    &   9 \\
Specific rule    & \multirow{2}{*}{1} & \multirow{2}{*}{1} & &  \multirow{2}{*}{1} & \multirow{2}{*}{4} & \multirow{2}{*}{5} & \multirow{2}{*}{3} & \multirow{2}{*}{2} & \multirow{2}{*}{17}\\
suggestion            & & & & & & & & & \\
Ask the      & \multirow{2}{*}{1} &    &    &    &  \multirow{2}{*}{2} &   \multirow{2}{*}{3} &    &   & \multirow{2}{*}{6}\\
decision subject               &    &    &    &    &    &    &    &  &  \\
Unclear*              &  2 &  3 &  2 &  7 &  4 &  6 &  2 & 5 & 31 \\
Not sure*             &    &  1 &    &  2 &  1 &    &    & & 4\\
\bottomrule     
\end{tabular}
\end{table*}

\subsection{Task-by-task preferences}
\begin{figure*}
    \begin{center}
            \includegraphics[width=1.0\textwidth, trim=0cm 3cm 0cm 3cm]{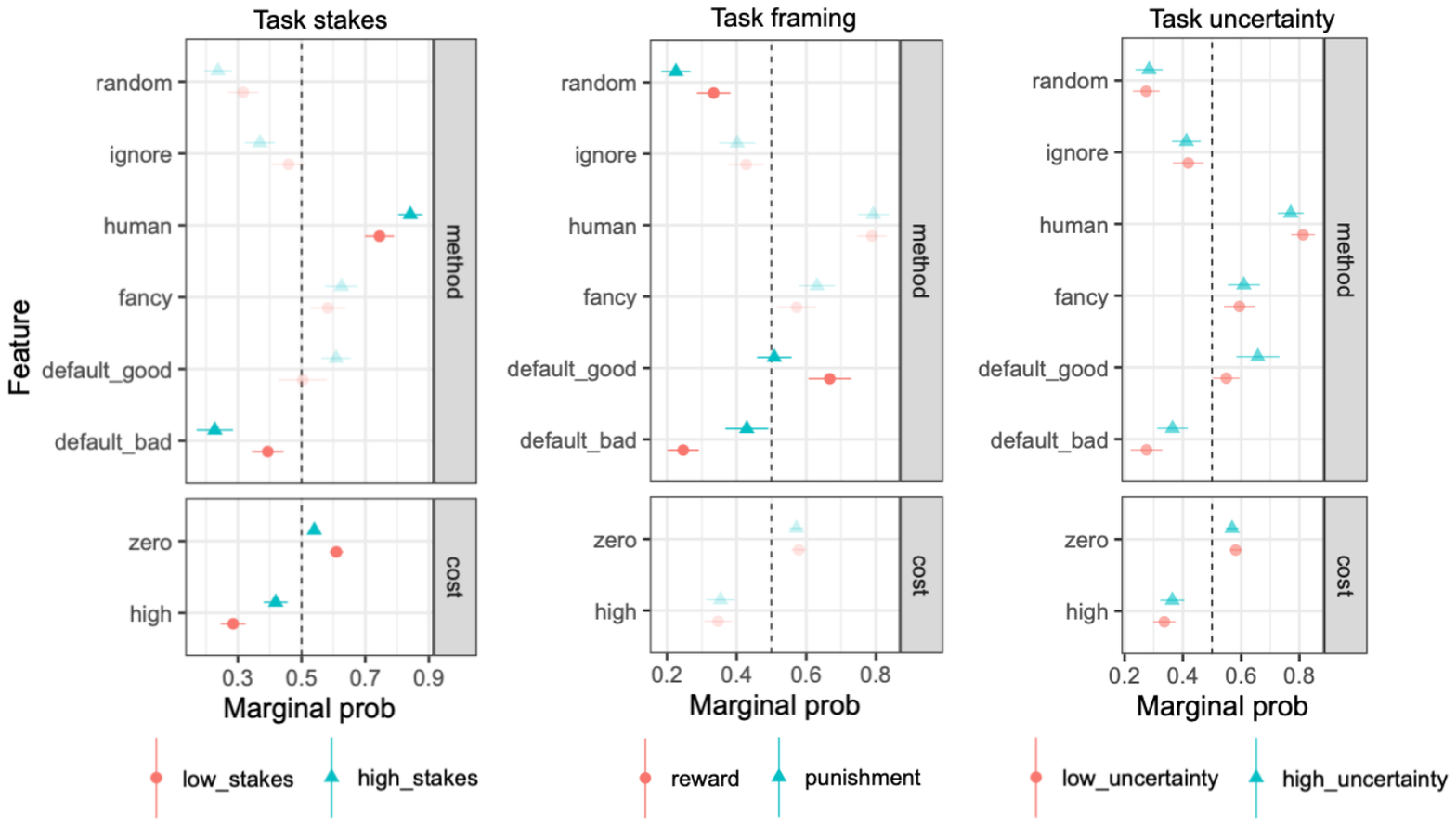}
    \end{center}
    \caption{ Marginal probabilities stratified by task stakes, task framing, and task uncertainty. For each ``feature'' (multiplicity resolution technique or cost level), the marginal probability is the chance of choosing that attribute, relative to a baseline random choice of 0.5. Results are further broken down by task characteristic. Results are darker when the confidence intervals for the two task levels do not overlap as a marker of greater significance. Confidences intervals were computed using $\alpha=0.05$ to ease readability, see the discussion in \cref{sec:res_hypo} for which differences are statistically significant with a multiple hypothesis correction. }
    \Description{Marginal mean preferences over multiplicity resolution techniques stratified by task stakes, task framing, and task uncertainty. We report which results have non-overlapping confidence intervals for each setting. For task stakes, human (more preferred in high stakes), default bad (more preferred in low stakes), zero cost (more preferred in low stakes) and non-zero cost (more preferred in high stakes). For task framing, random (more preferred for reward tasks), default good (more preferred for reward), and default bad (more preferred for punishment). For task uncertainty, no factor has non-overlapping confidence intervals.}
    \label{fig:part2_tasks}
\end{figure*}

\Cref{fig:part2_tasks} shows the relative preferences for each multiplicity resolution technique for tasks stratified by stakes, uncertainty, and framing. 

\subsection{Costs}
As noted in finding \explCost in \cref{sec:res_explore}, \nocost is significantly preferred over \extracost in aggregate. However, for three high-stakes settings (tumor detection, care plan access, and job probation), this distinction is not significant. The point estimates for marginal mean preferences for \extracost are 0.436 (SE 0.032, p<0.0001) for tumor detection, 0.474 (SE 0.038, p<0.0001) for care plan access, and  0.445 (SE 0.035, p<0.0001) for job probation. (A point estimate less than 0.50 indicates that \nocost is more preferred, and a point estimate of 0.50 indicates that there is no preferences between the cost options.) 

\subsection{Demographic factors}
There is no significant difference in preference for multiplicity resolution technique based on education level  or AI experience. That is, a t-test comparing the results for each resolution technique has a p-value greater than 0.05. When we consider age as a binary factor (divided at 35, roughly the median age and a common cutoff for ``young adult''), \texttt{fancy} is preferred by people younger than 35 (diff=0.081, SE=0.039, t= 2.054, p=0.040) and \texttt{default\_good} is preferred by people older than 35 (diff=-0.101, SE=0.040, t=-2.515, p=0.012). Looking at gender, \texttt{fancy} is preferred by men (diff=0.0783, SE=0.039, t=2.009, p=0.045) and \texttt{random} is preferred by women (diff=-0.070, SE=0.035, t=-2.027, p=0.043). We encourage future work to explore whether these trends hold as the experiment setup is varied.

\end{document}